# Origin of the Moon

Robin M. Canup[1], Kevin Righter[2], Nicolas Dauphas[3], Kaveh Pahlevan[4], Matija Ćuk[4], Simon J. Lock[5], Sarah T. Stewart[6], Julien Salmon[7], Raluca Rufu[7], Miki Nakajima[8], Tomáš Magna[9]

[1]*Planetary Sciences Directorate, Southwest Research Institute, 1050 Walnut Street, Boulder, CO 80302, U.S.A.*

[2]*NASA Lyndon B. Johnson Space Center Mail Code X12, 2101 NASA Parkway Houston, TX 77058, U.S.A.*

[3]*Department of the Geophysical Sciences and Enrico Fermi Institute, University of Chicago, 5734 South Ellis Avenue, Chicago, IL 60637, U.S.A.*

[4]*Carl Sagan Center for the Study of Life in the Universe, SETI Institute, 189 Bernardo Ave. Mountain View, CA 94043, U.S.A.*

[5]*Division of Geological and Planetary Sciences, California Institute of Technology, 1200 East California Boulevard, Pasadena, CA 91125, U.S.A.*

[6]*Earth and Planetary Sciences, University of California, Davis, One Shields Ave, Davis, CA 95616, U.S.A.*

[7]*Space Studies Department, Southwest Research Institute, 1050 Walnut Street, Boulder, Colorado, 80302 U.S.A.*

[8]*Department of Earth and Environmental Sciences, University of Rochester, 227 Hutchison Hall. Rochester, NY 14627, U.S.A*

[9]*Center for Lithospheric Research Czech Geological Survey Klárov 3 118 21 Prague 1, Czech Republic*

## 1. INTRODUCTION

The Earth-Moon system is unusual in several respects. The Moon is roughly ¼ the radius of the Earth – a larger satellite-to-planet size ratio than all known satellites other than Pluto's Charon. The Moon has a tiny core, perhaps with only ~ 1% of its mass, in contrast to Earth whose core contains nearly 30% of its mass. The Earth-Moon system has a high total angular momentum, implying a rapidly spinning Earth when the Moon formed. In addition, the early Moon was hot and at least partially molten with a deep magma ocean. Identification of a model for lunar origin that can satisfactorily explain all of these features has been the focus of decades of research.

*Apollo* and pre-*Apollo* era lunar origin theories included capture, fission, co-accretion, and collisional ejection (e.g., Wood 1986; Hartmann and Davis, 1975; Cameron and Ward 1976). Origin via a giant impact with the Earth emerged from the 1984 "*Origin of the Moon*" conference as the leading hypothesis, as it seemed best able to account for the Earth's rapid early spin, the Moon's small core, the Earth and Moon's similar oxygen isotopic compositions, and the Moon's hot start (Wood 1986). Impact simulations, initially computationally limited in number and resolution, demonstrated that a giant collision could produce an iron-poor, Earth-orbiting disk (Benz et al. 1986, 1987, 1989; Cameron and Benz 1991). Later models focused on identifying an impact that could also account for the Earth-Moon system angular momentum and the Moon's mass. One class of models required an impact when the Earth was only about half grown (Cameron 1997, 2000, 2001), which was difficult to reconcile with limited accretion of siderophile-rich material onto the Moon during the remainder of the Earth's growth. It was later shown that the impact of a Mars-sized body at the end of Earth's accretion could account for the main physical properties of the Earth-Moon system (Canup & Asphaug 2001; Canup 2004a,b), consistent with the impactor size originally suggested in Cameron and Ward (1976). However, this so-called "canonical impact" left a conundrum: it predicted the Moon formed primarily from material from the impactor, whereas the similarity in isotopic composition of the Earth and Moon (e.g., Wiechert et al. 2001; Zhang et al. 2012; Touboul et al. 2015; Kruijer et al. 2015) would be most easily explained if the Moon formed from the Earth. This quandary was recognized at the time of the prior *New Views of the Moon* book (e.g., Righter 2007; Taylor et al. 2006a), but has become more acute subsequently, and has been characterized by some as an "isotopic crisis" for the giant impact model (e.g., Melosh 2014).

Developments in laboratory techniques have produced copious new data relevant to lunar origin, and increasingly diverse impact scenarios have been proposed in an attempt to better reconcile impact models with both dynamical and geochemical constraints. The purpose of this chapter is to provide a broad overview of recent advances in these areas; interested readers are also referred to reviews by Asphaug (2014) and Barr (2016).

## 2. NEW DATASETS

We begin with a discussion of new measurements of the Moon's composition and how it compares to that of the bulk silicate Earth (BSE). Measured differences and similarities in chemical and isotopic compositions between terrestrial and lunar rocks bear on both the thermodynamic conditions of the Moon's accretion and the relative proportion of terrestrial vs. impactor material incorporated into the Moon as a function of the (uncertain) impactor composition. Both are key constraints for giant impact models.

### 2.1. Volatiles - H, C, S, N, Cl, F, alkalis, and volatile metals

The depletion in the abundances of volatile elements in the Moon relative to Earth's mantle was documented starting in the 1970s (Ringwood and Kesson 1976; Wolf et al. 1979; Wolf and Anders 1980). The Moon was also thought to be extremely dry relative to Earth (Taylor et al. 1995; Drake 2005), although bulk mantle values up to 600 ppm $H_2O$ were theoretically possible based on low pressure magma solubility limits (Abe et al. 2000; Righter and O'Brien 2011).

In the last decade, measurements of H, C, S, F, and D/H in lunar glasses (Saal et al. 2008; Hauri et al. 2011, 2015), apatites (Greenwood et al. 2011; Boyce et al. 2010, 2014; Barnes et al. 2013, 2014; Tartese et al. 2013), and nominally anhydrous minerals like plagioclase (Hui et al. 2013) challenged the long-held notion that the Moon is barren of volatile elements. Remote sensing has placed constraints on water abundances in magma sources ($M^3$ on Chandraayan-1; Pieters et al. 2009; Milliken and Li 2017; Klima et al. 2013) and in surficial deposits (LRO/LCROSS; Hurley et al. 2012; Colaprete et al. 2010). Bulk Cl contents and Cl isotope systematics have now been measured in *Apollo* samples, including mare basalts, pyroclastic glasses, soils, breccias, KREEP and highland rocks (Sharp et al. 2010; McCubbin et al. 2015).

Measurements of lunar glasses, apatites and nominally anhydrous minerals have been used to improve estimates of the composition of the accessible lunar mantle and bulk silicate Moon (BSM), i.e., the Moon's silicate composition just after accretion and before its differentiation into mantle and crust. The extent to which estimates are representative of the BSM is unknown as the Moon preserves stark chemical heterogeneities inherited from lunar magma ocean (LMO) crystallization. With this caveat in mind, estimates of volatiles in the BSM are ~ 3 to 292 ppm $H_2O$, 14 to 570 ppb C, 75 to 80 ppm S, 0.145 to 129 ppm Cl, and 4.5 to 60 ppm F (see review by McCubbin et al. 2015; Hauri et al. 2015; Chen et al. 2015). For comparison, the BSE estimates of these elements are ~1200 ppm ($H_2O$), 100 ppm C, 200 ppm S, 30 ppm Cl, and 25 ppm F (Palme and O'Neill 2014; Dauphas and Morbidelli 2014; Marty 2012; Halliday 2013). For some elements these are the first robust estimates for the BSM, and for others such as S, the estimates are similar to previous studies (e.g., Bombardieri et al. 2005; 75 ppm S). Surprisingly, the new data imply that at least some portions of the early Moon were water-bearing, with $H_2O$ abundances comparable to those estimated for the Earth's mantle (e.g., Greenwood et al. 2011; Hauri et al. 2011; McCubbin et al. 2010). The data also suggest that the BSM was depleted compared with the BSE by a factor of 2-3 for S and >100 for C, whereas the BSM appears comparable to the BSE for Cl and F. The sulfur isotopic homogeneity of lunar mare basalts indicates that there has been loss of only < 1 to 10% of lunar S following lunar accretion (Wing and Farquhar 2015). Although S is depleted in the BSM, like water its abundance is anomalously elevated given its higher volatility than alkali elements in the solar nebula, a circumstance that has previously been used to argue for a low hydrogen fugacity for lunar volatile depletion (Ringwood and Kesson 1977).

Other volatile elements are chalcophile or siderophile, such as Ga, Ge, Cu, Zn, Sn, Cd, In, Bi, Ag, but their partitioning into a small core (with ~ 1.5% of the Moon's mass; Williams et al. 2014) was understood at only a reconnaissance level in the 1970s (Ringwood and Kesson 1976), making models for their partitioning into the lunar core and associated depletion somewhat qualitative. Experimental studies have now established how these elements partition between metal and silicate as a function of pressure, temperature, oxygen fugacity, and the composition of the metallic (i.e., mole fraction of C and S in the lunar core) and silicate liquids (e.g., Mann et al. 2009; Ballhaus et al. 2013; Wang et al. 2016; Righter et al. 2017b). New elemental and phase equilibria data, coupled with re-analysis of seismic data, moment of inertia, and tidal Love number, can constrain the Moon's interior structure (e.g., Righter et al. 2017a; Weber et al. 2011; Williams et al. 2014; Khan et al. 2014). Uncertainties in the density of the core, however, still allow a range of lunar core compositions, from nearly C- and S- free at the high end (e.g., Righter et al. 2017a) to S- or C-bearing (e.g., up to 8% S; Antonangeli et al. 2015) at the low end. The early Moon is thought to have been largely molten, and for expected core-mantle boundary pressures and oxygen fugacities (e.g., Karner et al. 2006; Righter et al. 2018a; Steenstra et al. 2016), the lunar Ga, Ge,

Cu, Ag, and P contents can be explained if an already volatile-depleted BSM (by a factor of 4 to 5 relative to BSE) equilibrated with a small lunar core (Righter 2019).

[Figure 1]

Elemental abundances in the Earth and Moon compared to those in CI chondrites are summarized in Figure 1. Most refractory lithophile and weakly siderophile elements are similar in the BSM and BSE, but FeO appears significantly higher in the lunar mantle (e.g., Dauphas et al. 2014a and references therein), an important (and somewhat overlooked) constraint for models of lunar origin. In contrast, the volatile alkali elements K, Na, Rb, and Cs are depleted by a factor of 5-6 in the BSM relative to the BSE, sometimes illustrated by the lower K/U (wt. ratio) in the Moon, ~2500, relative to Earth, ~12000 (e.g., Taylor 1982; McDonough et al. 1992). In addition, S is depleted by a factor of 2-3, and C and other moderately and highly volatile siderophile elements are depleted by factors of 5 to 200 relative to the BSE; concentrations of these two elements were established in the BSE by a combination of core formation and late chondritic additions (or late veneer) (e.g., Boujibar et al. 2014; Li et al. 2015). Lithium has a higher condensation temperature (~1150 K; Lodders 2003) than the other alkalis and the BSM and BSE have similar Li concentrations, indicating no relative depletion in the Moon. Lunar concentrations of the moderately volatile siderophile elements (in order of increasing volatility As, Cu, Ag, Sb, Ga, and Ge) could be explained by a combination of starting material that is Earth-like in its volatile depletion, followed by formation of a small lunar core, but the highly volatile elements (Bi, Cd, Sn, In, Zn, Tl, Pb, and possibly C) are depleted in the Moon to a greater extent than expected by this combination, and thus require either an initial greater depletion in these elements or an additional volatile loss mechanism (Righter 2019).

The mechanism responsible for the Moon's volatile depletion relative to that of the Earth is still debated (e.g., Day and Moynier 2014; see also §4.5). It could simply reflect an initially volatile-poor composition for the protolunar material, perhaps due to incorporation of substantial material from a volatile-depleted impactor (e.g., Newsom and Taylor 1989), but this could account for only a factor of ~5-6 depletion (at most) and would not explain the higher depletions documented for Zn and other elements. Alternatively, the protolunar material could have initially been approximately Earth-like in its volatile inventory, with subsequent depletion associated with escape from the hot protolunar disk material (Genda and Abe 2003; Desch and Taylor 2011), melt-vapor fractionation in the disk leading to the preferential incorporation of the melt phase in the Moon (e.g., Canup et al. 2015; Charnoz and Michaut 2015; Pahlevan et al. 2016; Lock et al. 2018; Nie and Dauphas 2019), and/or much later degassing and loss from a lunar magma ocean or other magmatic activity (e.g., Elkins-Tanton and Grove 2011a, Boyce et al. 2015; Barnes et al. 2016; Steenstra et al. 2017; Righter et al. 2018a; Dhaliwal et al. 2018). Also, the extent of volatile loss across planetary bodies could be associated with basic planetary parameters, such as surface gravity, escape velocity and mass (Day and Moynier 2014), meaning that the retention of volatiles acquired during accretion may be a function of planet size. Whatever mechanism(s) occurred appears to have preferentially affected elements of higher volatility (Righter 2019). Isotopic variations in Zn, K, Rb, Cu, Ga, and Sn (§2.4) may help distinguish between competing volatile depletion hypotheses.

The exciting finding of higher volatile contents than expected in some lunar materials has stirred the field of lunar geochemistry. However, further work constraining the water, Cl and F contents of deeply formed lunar materials is needed to resolve the question of how water-rich and volatile-rich the bulk lunar interior was compared with the Moon's generally very dry surface samples, and how the Moon's initial volatile content compared with that of the BSE (e.g., Robinson and Taylor 2014; Robinson et al. 2016; Cano et al. 2020; see also McCubbin et al. 20XX, this volume).

## 2.2. Importance of isotopes

Most geological processes such as melting, magmatic differentiation, and aqueous alteration produce predictable mass-dependent isotopic variations, meaning that the variations scale as the difference in mass of the isotopes involved. These mass-dependent isotopic variations can provide clues on the processes that established the chemical composition of planets.

Observed isotopic variations that depart from mass-dependent fractionation and are not due to radiogenic and/or cosmogenic effects (e.g., §2.5) are called isotopic anomalies. In most cases, isotopic anomalies reflect differences inherited from the protoplanetary disk (e.g., Dauphas et al. 2002; Clayton 2002), developed through incomplete mixing of the products of nucleosynthesis in the solar nebula, or via exotic photochemistry in molecular clouds or the early solar system (Dauphas and Schauble 2016 and references therein). Planetary processes can alter an element's absolute concentration, but can hardly affect its isotopic anomalies. This makes isotopic anomalies ideal tracers to establish the parentage of planetary bodies, and they have provided the strongest constraints on the fraction of the Moon that originated from the Earth, or alternatively, the degree to which the Moon-forming giant impactor was Earth-like in its composition.

Isotopic anomalies are usually reported by calculating the departure of an isotopic ratio from that of a reference material after correcting for mass-dependent fractionation by normalizing one ratio to a fixed value. A minimum of 2 isotopic ratios (or 3 isotopes) is needed to define a mass-dependent fractionation line and associated isotopic anomalies. Consider oxygen as an example, which has 3 isotopes, $^{16}O$, $^{17}O$, and $^{18}O$, such that 2 isotopic ratios can be defined, $^{17}O/^{16}O$ and $^{18}O/^{16}O$ (Clayton et al. 1973). Isotopic ratios are often expressed in δ-notation, which is the departure in parts per 1,000 (permil, or ‰) of the measured ratio from that of a reference reservoir (e.g., V-SMOW or Vienna Standard Mean Ocean Water for oxygen), with $\delta^{17,18}O\ (‰) = [(^{17,18}O/^{16}O)_{Sample}/(^{17,18}O/^{16}O)_{Standard} - 1] \times 1000$. Because the difference in mass between $^{17}O$ and $^{16}O$ is half as large as the mass difference between $^{18}O$ and $^{16}O$, most terrestrial samples plotted in a $\delta^{17}O$ vs. $\delta^{18}O$ diagram fall on a line with an approximate slope (17−16)/(18−16)=1/2 that passes approximately through the [0,0] coordinate at which the sample's oxygen isotopic values equal that of SMOW. Isotopic anomalies in planetary materials are defined as vertical offsets from this line and are expressed in parts per thousand units as $\Delta^{17}O$ values. The same approach can be extended to systems of more isotopes for elements heavier than oxygen. The main difference is that these systems often show more subtle isotopic variations, and the departures from a standard are therefore reported using ε-notation, which is the deviation in parts per 10,000.

In the following sections, we examine the isotopic composition of the Moon relative to the Earth. To be meaningful, such discussion must be considered within the context of isotopic variations observed in meteoritic materials originating from the asteroid belt and Mars. For some elements, all meteorites have the same isotopic composition to within current analytical precision, and no isotopic differences between the Earth and Moon are expected or found. Other elements such as Zr show only subtle isotopic variations among meteoritic materials (relative to measurement error bars) and their composition provides minimal constraints on scenarios of lunar formation (Akram and Schönbächler 2016).

We first focus on O, H, Ti, Ca, and Cr as elements that display a wide range of isotopic anomalies and that have been precisely measured in lunar and terrestrial samples (§2.3). We then discuss observed mass-dependent isotopic fractionations that can provide clues on the mechanisms of volatile loss (K, Rb, Cu, Ga, Zn, and Sn) or core formation (Si, V, and Fe) (§2.4). Finally we discuss isotopic variations that are radiogenic in origin, including those of W (§2.5).

## 2.3 Distinctive isotopic systems (O, H, Ti, Ca, and Cr)

Following the discovery of oxygen isotopic anomalies in carbonaceous chondrites (Clayton et al. 1973), Clayton and Mayeda (1975) first showed that the oxygen isotopic compositions of *Apollo* samples fall on the same mass fractionation line as terrestrial rocks, meaning that no other source reservoir was detected within measurement uncertainties. The mean $\Delta^{17}O$ value calculated from their data was $-0.075\pm0.049$ ‰. Later work by Clayton and Mayeda (1996) revealed a $\Delta^{17}O$ value of $+0.01\pm0.06$ ‰ for lunar meteorites, compared to $-0.02\pm0.05$ ‰ for terrestrial tektites. These relatively large error bars by modern standards allowed for a small but significant isotopic difference between the Earth and Moon within uncertainties. A clearly resolved isotopic offset of SNC meteorites derived from Mars was reported as $\Delta^{17}O = +0.321\pm0.013$‰ by Franchi et al. (1999), which suggested substantial isotopic variation among planetary bodies during accretion.

Subsequent giant impact simulations defined a "canonical" Moon-forming impact and found that the protolunar material originates primarily from the impacting planet, rather than from the protoearth (Canup and Asphaug 2001; Canup 2004a,b, 2008; Canup et al. 2013). If the impactor had a different isotopic composition than the protoearth, one would then expect to see Earth-Moon differences today. This has motivated increasingly precise O isotope measurements to determine whether such a difference exists. Wiechert et al. (2001) obtained a mean $\Delta^{17}O = +0.003\pm0.005$ ‰ for lunar rocks, or equivalently $\Delta^{17}O = +3\pm5$ parts per million (ppm), meaning that the lunar samples are indistinguishable from the terrestrial $\Delta^{17}O$ value (0 ‰, by definition). Spicuzza et al. (2007) and Hallis et al. (2010) subsequently reported $\Delta^{17}O = +0.008\pm0.010$ ‰ (8 $\pm10$ ppm) and $-0.008\pm0.021$ ‰ ($-8\pm21$ ppm), respectively, for lunar rocks, supporting the view that Earth and Moon have the same O isotopic compositions. This view was questioned by Herwartz et al. (2014), who measured a difference of $+0.012\pm0.003$‰ ($12\pm3$ ppm) between three lunar samples and terrestrial rocks. Young et al. (2016) evaluated a set of 14 lunar samples and several terrestrial samples and obtained an Earth-Moon $\Delta^{17}O$ difference of $-0.001\pm0.005$‰ ($-1\pm5$ ppm), reinstating the Wiechert et al. (2001) conclusion that lunar and terrestrial rocks had identical O isotopic compositions. Greenwood et al. (2018) measured a large set of lunar (17 whole rocks and 14 mineral separates) and terrestrial samples and found a statistically significant difference of 3 to 4 ppm between the two. Cano et al. (2020) emphasized that oxygen isotopic compositions vary with lunar lithology (Figure 2A), and proposed that samples with a deep source, notably VLT green glasses, are most indicative of the Moon's starting composition. Based on those samples, Cano et al. argued for an initial Earth-Moon difference of ~20 ppm or more (Figure 2A).

Future data may better constrain the relative Earth-Moon oxygen isotopic compositions. However, avoiding biases is difficult at this level of precision; for example, even differences between the assumed slope associated with mass-dependent fractionation (which is ≈ 0.5 but can be as large as 0.53 for high-temperature, igneous processes) can be important (e.g., Young et al. 2016; Sharp et al. 2018). Nevertheless, a robust conclusion is that the Earth and shallow Moon have $\Delta^{17}O$ values within ~0.01‰ (10 ppm) of each other, whereas samples derived from the deeper lunar interior exhibit differences up to about 20 ppm (Fig. 2A). In contrast, common meteorites span a range of $\Delta^{17}O$ values from −4.7 ‰ (-4700 ppm; Eagle Station Pallasite) to +2.6 ‰ (+2600 ppm; R chondrites), a factor of several hundred times larger range than the maximum inferred Earth-Moon difference (Dauphas and Schauble 2016; Dauphas 2017).

One possible explanation for the extreme closeness in the Earth-Moon O isotopic compositions is an Earth-like composition impactor, presumably related to the family of enstatite meteorites. Although enstatite chondrites display a broad range in $\Delta^{17}O$ (roughly −300 ppm to + 350 ppm; e.g., Greenwood et al. 2018), their mean values are similar to those of the Earth and Moon. Newton et al. (2000) found $\Delta^{17}O$ = +0.010±0.12‰ for EH chondrites and $\Delta^{17}O$ = +0.014±0.06‰ for EL chondrites; a more recent compilation in Dauphas (2017) gives $\Delta^{17}O$ = −0.03±0.10‰ and −0.01±0.07 ‰ for EH and EL, respectively. The aubrite class of enstatites, thought to reflect a differentiated parent body, also have relatively similar $\Delta^{17}O$ values to those of lunar and terrestrial rocks, with $\Delta^{17}O$ = +0.028±0.003‰ (based on data from Barrat et al. 2016, recalibrated in Greenwood et al. 2018). A recent study of enstatite-rich lithologies in the Almahata Sitta ureilite suggest these originated from a 500 km-sized differentiated parent body of enstatite chondrite parentage (Harries and Bischoff 2020). Such material is poorly represented in meteorite collections, suggesting that such collections may provide an incomplete and/or biased representation of enstatite related materials.

Hydrogen isotopes have been measured in lunar basalt, volcanic glasses, apatites, and magmatic melt inclusions, resulting in a wide range of deuterium-to-hydrogen (D/H) ratios. This range could be generated by input to the Earth-Moon system from several sources, including solar nebular (Hallis et al. 2015), cometary (Greenwood et al. 2011), and chondritic (Tartese and Anand 2013; Füri et al. 2014; Piani et al. 2020). D/H measurements have been used to argue that inner solar system hydrogen (or water) sources are very similar and that the Moon acquired its low water content from the same source as Earth (Saal et al. 2013; Sarafian et al. 2017); some have argued that the low D/H on Earth and Moon is similar to that found in enstatite chondrites (Piani et al. 2020). The interpretation of sources of lunar hydrogen isotopic values is complicated by later thermal, impact, exposure, and magmatic processes, all of which can increase or decrease D/H in lunar materials (e.g., Treiman et al. 2016; Robinson et al. 2016). Thus, hydrogen isotopes are a promising tool for unravelling the origin of lunar volatiles, but ongoing debate demonstrates that additional work is necessary for a comprehensive understanding (see also McCubbin et al. 202X, this volume).

[Figure 2]

Returning to oxygen, it has been argued that the Earth–Moon similarity in O isotopes could reflect isotopic equilibration between the terrestrial magma ocean and protolunar disk via the vapor phase (Pahlevan and Stevenson 2007). Depending on the conditions during equilibration (§4.2), this process may preferentially affect elements in the vapor phase, and could be less effective for highly refractory elements. A potential test of the equilibration hypothesis is then to compare the isotopic compositions of lunar and terrestrial rocks in refractory elements that display isotopic variations among meteorites (§5.2). Titanium is refractory and displays isotopic anomalies involving several isotopes that are correlated in meteorites, with the most prominent being the neutron-rich isotope $^{50}$Ti, whose anomalies expressed as $\varepsilon^{50}$Ti range from −1.9 in ureilites to +3.8 in CO chondrites (Trinquier et al. 2009; Zhang et al. 2011, 2012). Zhang et al. (2012) measured the Ti isotopic composition of *Apollo* lunar samples and found variations in $\varepsilon^{50}$Ti values between −0.23±0.04 and +0.02±0.05. These variations correlate with $^{150}$Sm/$^{152}$Sm and $^{158}$Gd/$^{150}$Gd ratios, which are proxies of cosmogenic neutron capture effects, indicating that in some lunar samples the Ti isotopic composition was modified by exposure to cosmic rays. Using the correlations with Sm and Gd isotopes to correct for these effects, Zhang et al. (2012) estimated a pre-irradiation bulk lunar $\varepsilon^{50}$Ti value of −0.03±0.04 (Figure 2B). Terrestrial rocks have by definition $\varepsilon^{50}$Ti = 0, meaning that no $^{50}$Ti anomaly was detected in lunar rocks. Zhang et al. (2012) also estimated the equilibration timescale (i.e., the timescale for exchange between the magma disk and the vapor atmosphere) for Ti, and despite the low vapor pressure of this element relative to oxygen, exchange between liquid and vapor could have taken place within the lifetime of the disk, meaning that its composition could have been affected by equilibration.

Meteorites also display isotopic anomalies in $^{48}$Ca that correlate tightly with $^{50}$Ti isotopic anomalies (Dauphas et al. 2014b). Variations in $\varepsilon^{48}$Ca among meteorites range from −1.8 to +3.9, similar to $\varepsilon^{50}$Ti because the two follow a 1:1 correlation. Calcium data are, however, not directly comparable with $^{50}$Ti because Ca is more refractory ($T_c(50\%) = 1659$ K, compared to 1593 K for Ti; Lodders 2003) with a longer (~30-yr) timescale for liquid-vapor equilibration under expected protolunar disk conditions (Zhang et al. 2012). Early measurements of lunar samples gave an $\varepsilon^{48}$Ca value of −0.24±0.24 (Dauphas et al. 2015a), implying that either identical Earth-Moon Ca isotopic compositions or a difference of ~ −0.5 $\varepsilon$ units was equally likely. With improved precision, Schiller et al. (2018) report $\varepsilon^{48}$Ca = 0.0020±0.039 for terrestrial rocks and $\varepsilon^{48}$Ca = 0.037±0.019 for 4 lunar meteorites (weighted averages and 95% confidence intervals). These are consistent with a small, < 10 ppm Earth-Moon difference, but equal compositions are also possible within the uncertainties. More work is needed to scrutinize the relative Earth-Moon calcium isotopic compositions, because preservation of a heterogeneity in highly refractory Ca may be expected for vapor-mediated equilibration (Zhang et al. 2012).

Chromium also displays variations in its neutron-rich isotopes, with $\varepsilon^{54}$Cr values ranging from −0.9 (ureilites) to +1.6 (CI chondrites) (Trinquier et al. 2007; Qin et al. 2010). A difficulty with Cr in lunar samples is that its isotopic composition has been modified by exposure to cosmic rays through the spallation (break-up) of iron atoms, with this effect expected to increase the $\varepsilon^{54}$Cr value of lunar rocks over time. Taking the lowest of the $\varepsilon^{54}$Cr values measured by Qin et al. (2010) in two lunar samples as an upper limit on the pre-exposure lunar $\varepsilon^{54}$Cr value gives +0.08±0.12, indistinguishable from the terrestrial composition within uncertainty. Mougel et al. (2018) attempted to quantify and correct for spallogenic effects by correlating Cr isotopic shifts with $^{150}$Sm/$^{152}$Sm ratios. An issue with this approach is that Sm isotopes are affected by capture of secondary neutrons, while excess $^{54}$Cr is produced by iron spallation, and the two processes occur at different depths in the target sample and have different sensitivities to bulk-rock chemistry ($^{54}$Cr production is highly sensitive to the Fe/Cr ratio of the target material). With this caveat in mind, the correlations between $^{150}$Sm/$^{152}$Sm and $\varepsilon^{54}$Cr give a pre-exposure $\varepsilon^{54}$Cr value of +0.09±0.08, which is almost identical to terrestrial composition but would still allow for the presence of a potentially significant isotopic anomaly in Cr.

To summarize, all refractory (non-volatile) elements that are known to display isotopic anomalies in meteorites and that have been measured in lunar rocks show isotopic compositions that are essentially indistinguishable from (or very similar to) terrestrial composition within current uncertainties. Our Moon is distinctively Earth-like in its isotopic anomaly composition across multiple key elements, a profoundly important constraint for lunar origin models. Determining whether the small differences from terrestrial isotopic values (e.g., Ca and Cr) measured in some lunar samples are due to heterogeneity within the Moon, impactor contributions, and/or incomplete Earth-Moon equilibration remains a goal of future isotopic studies.

## 2.4 Isotopes that are volatile K, Rb, Cu, Zn, Ga, Sn, and/or may participate in core formation (Si, Fe, V)

Isotopic fractionations have been documented in lunar rocks for many moderately volatile elements, notably K (Wang and Jacobsen 2016; Tian et al. 2020), Rb (Pringle and Moynier 2017; Nie and Dauphas 2019), Cu (Herzog et al. 2009), Zn (Paniello et al. 2012; Kato et al. 2015), Ga (Kato and Moynier 2017), and Sn (Wang et al. 2019). In particular, lunar rocks were shown to be enriched in the heavy isotopes of K ($\delta^{41}$K) and Rb ($\delta^{87}$Rb) by ~+0.4 and +0.16 ‰, respectively (Wang and Jacobsen 2016; Pringle and Moynier 2017; Nie and Dauphas 2019). As discussed by Nie and Dauphas (2019), equilibrium isotopic fractionation between vapor and liquid could shift the K isotope composition of the residual melt in a Rayleigh distillation by at most ~+0.03 ‰ at 3500 K to ~+0.15 ‰ at 1650 K, which is insufficient to explain the heavy K isotope enrichment measured in lunar rocks. These enrichments most likely reflect kinetic isotope effects associated with evaporation. Sossi et al. (2018) argued that Cr isotopic fractionation measured in lunar rocks required loss of Cr by equilibrium evaporation at low temperature (1700 K) and under highly oxidizing conditions. Shen *et al*. (2020), however, showed that the isotopic fractionation measured by Sossi *et al*. (2018) most likely reflects stable isotopic fractionation during magmatic differentiation, with the Earth and Moon having identical $\delta^{53}$Cr values. Wang et al. (2019) found that lunar rocks have light $\delta^{124/116}$Sn isotopic composition relative to Earth, which at face value cannot be explained by kinetic isotopic fractionation during evaporation as this process can only produce heavy isotope enrichments in the residue. Wang et al. (2019) argued that these light isotope enrichments could be explained by equilibrium fractionation between vapor SnO and Sn dissolved in silicate melt. As with K, it is not clear whether equilibrium can quantitatively explain the isotopic composition of lunar rocks for Sn, even at the lowest temperatures considered.

Kinetic effects associated with differences in the rates of evaporation of different isotopes can create large isotopic fractionation (Richter et al. 2002, 2009; Richter 2004), even at the extreme temperatures considered in the aftermath of the Moon-forming impact. For example, 80% loss of K under vacuum conditions would enrich the residue by ~+34‰ (Richter et al. 2011). Lunar rocks show heavy K isotope enrichments, but of much more limited magnitude, meaning that volatile loss occurred under conditions that limited the expression of kinetic isotope effects. Several processes can affect kinetic isotopic fractionation during evaporation, notably transport in the melt, in the vapor, and evaporation in a closed system (*e.g.*, Richter et al. 2002, 2007, 2009, 2011; Richter 2004; Dauphas et al. 2015b; Wang et al. 1999; Young et al. 1998, 2019; Ozawa and Nagahara 2001; Bourdon and Fitoussi 2020; Yamada et al. 2006; Galy et al. 2000).

Nie and Dauphas (2019) proposed a model for the depletion of moderately volatile elements in the Moon that can potentially explain the heavy isotope enrichments measured for most MVEs. They proposed a specific setting where this could be achieved, which is removal of the vapor disk by viscous drainage onto the Earth (Charnoz and Michaut 2015). Under those conditions and provided that the evaporation coefficients of the different MVEs are not too different, one would expect the vapor to be characterized by a similar undersaturation $S = P/P_{eq}$ for all elements. If the vapor is removed rapidly, $S \simeq 0$ and large kinetic isotope fractionation is produced (like vacuum evaporation), which is not seen. If the vapor is removed slowly, $S \simeq 1$ and the isotopic fractionation is close to equilibrium, which is too small to explain the data. The K and Rb data can be explained by Rayleigh evaporation under a vapor saturation of 0.99.

Measurements of lunar rocks show that they are also characterized by heavy Cu, Zn, and Ga isotopic compositions (Herzog et al. 2009; Kato et al. 2015; Kato and Moynier 2017; Wimpenny et al. 2018, 2019), although different lunar lithologies can show significant isotopic variations and defining a precise lunar isotopic composition is difficult. Part of that variability could stem from vapor loss during and after the stage of LMO (Kato et al. 2015; Kato and Moynier 2017; Day et al. 2017, 2020; Tian et al. 2020; Wimpenny et al. 2020; Dhaliwal et al. 2018). Taking the current best estimated Cu, Ga, and Zn isotopic compositions for the BSM, Nie and Dauphas (2019) showed that the differences in isotopic compositions for Rb, Ga, Cu, K, and Zn between BSM and BSE correlate with the quantity $\Delta_{\text{kin}} \ln f_{\text{BSM}}$ (Fig. 2C), where $\Delta_{\text{kin}} = 1000 \left[ (m_j/m_i)^\beta - 1 \right]$ is the kinetic isotopic fractionation between vapor and liquid associated with free evaporation and $f_{\text{BSM}}$ is the degree of elemental depletion of BSM relative to BSE for the element considered. Such a linear relationship is expected if all elements had similar evaporation coefficients and they were lost to a medium that was 98-99% saturated. This 1-2% departure from equilibrium is sufficient to explain the heavy isotope enrichments documented for most MVEs in lunar rocks. Wang et al. (2019) measured light Sn isotopic composition in lunar rocks, which cannot be explained by kinetic isotopic fractionation during evaporation (Figure 2C). They argued that this peculiar behavior reflected equilibrium fractionation between SnO in the vapor and Sn dissolved in silicate liquid. This observation can potentially be reconciled with the isotopic measurements made on other MVEs if the evaporation coefficient of Sn is higher than those of other MVEs, which would allow faster vapor-liquid equilibration. This is an ad hoc explanation and more work is needed to measure or calculate evaporation coefficients of MVEs under lunar formation conditions, and evaluate whether the Sn isotopic composition measured by Wang et al. (2019) is representative of the bulk Moon.

If most MVEs were lost by evaporation under 99% saturation, this provides constraints on rates and timescales, as the rate of evaporation of a volatile species is controlled by its saturation level through the Hertz-Knudsen equation. In the framework of the Charnoz and Michaut (2015) model, Nie and Dauphas (2019) showed that the gas viscosity required to explain the 99% saturation of MVEs was consistent with the timescale of vapor drainage onto Earth driven by magnetorotational instability. Other scenarios of MVE depletion (Canup et al. 2015; Lock et al. 2018) may be able to achieve evaporation under a similar vapor saturation, but this has not yet been carefully evaluated.

Silicon isotopic systematics may also help constrain lunar formation theories. Available data for lunar rocks cluster at $\delta^{30}\text{Si} = -0.29 \pm 0.03$‰, indistinguishable from Earth's mantle $\delta^{30}\text{Si}$ value of −0.30 ± 0.03‰ (Armytage et al. 2012; Poitrasson and Zambardi 2015). Si isotopes can be fractionated by core formation (e.g., Shahar et al. 2011; Hin et al. 2014; Moynier et al. 2020) and post-giant-impact evaporation-condensation processes (e.g., Pahlevan et al. 2011), but Si isotopic differences may also exist in pre-cursor nebular materials (Pringle et al. 2014; Dauphas et al. 2014a). The Si isotopic composition of the BSE is heavy compared with chondrites, which have $\delta^{30}$Si from approximately −0.7‰ to −0.35‰. It has been considered that this may reflect equilibrium Si isotopic fractionation between metal and silicate during terrestrial core formation, with experimental and theoretical evidence suggesting that light Si isotopes preferentially enter metal phases and, by inference, the core (e.g., Georg et al. 2007; Shahar et al. 2011; Hin et al. 2014). If the Moon was made from material derived from a smaller impactor than Earth, the Si isotopic composition would likely be different in the two mantles owing to the different PT conditions of core formation (e.g., Armytage et al. 2012). The difficulty with this interpretation is that terrestrial core formation took place at high temperatures (~3750 K; e.g., Righter et al. 2018b) where little isotopic fractionation is predicted between silicate and metal. Because most Si is in the mantle, there is limited leverage to change its mantle isotopic composition. Hin et al. (2014) concluded that unrealistically high amounts of Si in Earth's core would be needed to explain the difference in Si isotopic composition between terrestrial mantle rocks and chondrites. An alternative explanation is that this difference in $\delta^{30}$Si and Mg/Si ratio is not a signature of core formation but was instead inherited in part from impacts among precursors to the Earth or nebular processes (Pringle et al. 2014; Dauphas et al. 2014a). Indeed, the angrite meteorites display a heavy isotopic composition similar to that of the Earth and Moon that could not have originated via core formation (Pringle et al. 2014; Dauphas et al. 2014a).

The implications of the identical Earth-Moon Si isotopic composition for models of lunar origin thus remain ambiguous: it could reflect a Moon formed from material that equilibrated with the terrestrial mantle, or it could reflect a protoearth and impactor whose mantles had similar heavy Si compositions due to non-core formation processes, as is seen in angrites. In either case, the absence of Si isotopic fractionation between the silicate Earth and Moon constrains liquid-vapor fractionation and equilibration during the lunar formation process (e.g., Pahlevan et al. 2011; see §4.2.2 and §5.2).

Both Fe and V are redox sensitive elements and have potential to record isotopic variations during core formation and metal-silicate equilibrium, and may also provide information about planetary building blocks (e.g., Poitrasson et al. 2004; 2019; Hopkins et al. 2019). Similar, but not identical, V isotopic values for the bulk Moon (corrected for cosmogenic exposure as with Ti) and the BSE were demonstrated by Hopkins et al. (2019), indicating similarity to Earth, but slight variations within chondrite groups and other achondrite bodies could be due to nucleosynthetic effects, evaporative effects, or metal-silicate equilibrium (e.g., Nielsen et al. 2019, 2020). Similarly, for Fe isotopes, the isotopic composition of the BSE is debated and there remains uncertainty over whether core formation can fractionate iron isotopes (e.g., Elardo and Shahar 2017). Lunar mare basalts also show Fe isotopic variations that are difficult to explain (Poitrasson et al. 2019). In general, V and Fe isotopes may hold important information about the Earth-Moon system, but some key information is still lacking.

**2.5. Radiogenic isotopes – Hf-W, Sm-Nd, and Lu-Hf: Constraints on Earth-Moon relationships and formation chronology**

At the time of *New Views of the Moon* (2006), lunar chronology was dominated by Hf-W isotopic systematics and Sm-Nd FAN ages; the former indicated early differentiation (and thus formation) of the Moon (e.g., Kleine et al. 2009), while the latter suggested slightly younger ages, perhaps due to thermal resetting or a prolonged magma ocean stage. New constraints have come from Hf-W measurements on a wide range of lunar samples with improved corrections for $^{182}$W originating from cosmic ray produced $^{182}$Ta, Sm-Nd dating of lunar anorthosites, and Lu-Hf and U-Pb dating of zircons from *Apollo* 14 breccia and soil samples. While isotopic variations in $^{182}$W were initially sought to constrain the age of the Moon, they have played an increasingly important role in constraining the origin of the material that made the Moon (e.g., Kruijer et al. 2015; Touboul et al. 2015; Kruijer and Kleine 2017).

***2.5.1 Constraints from the Hf-W system.*** Tungsten displays small isotopic anomalies in bulk meteoritic materials (Qin et al. 2008) but the focus in Earth-Moon comparisons has been on the radiogenic isotope $^{182}$W, which is produced by decay of the short-lived radionuclide $^{182}$Hf whose half-life is $t_{1/2}$=8.9 Myr (Kleine et al. 2009; Vockenhuber et al. 2004). Determination of the W isotopic composition of the Moon has a long history (e.g., Lee et al. 2002; Kleine et al. 2002) that for sake of space will not be repeated here. The crux of the issue is that the abundance of $^{182}$W on the lunar surface is due not only to the radiogenic decay of $^{182}$Hf, but also to modification by cosmic ray exposure, which produces $^{182}$W via irradiation of $^{181}$Ta and removes $^{182}$W by neutron-capture production of $^{183}$W (Leya et al. 2000). Various approaches have been used to account for cosmic ray effects to determine the Moon's pre-exposure, indigenous W isotopic composition. Touboul et al. (2015) considered Ta-free lunar metals and obtained an indigenous lunar $\varepsilon^{182}$W value of +0.206±0.051 relative to the current terrestrial mantle (with $\varepsilon^{182}W_\oplus = 0$ by definition). Kruijer et al. (2015) and Kruijer and Kleine (2017) used measurements of Hf isotopes to correct for neutron capture effects and obtained $\varepsilon^{182}W = +0.27\pm0.04$.

Lunar mantle reservoirs produced during magma ocean solidification have varied Hf/W compositions (e.g., the source of high-Ti mare basalts vs. KREEP), and if solidification occurred during the lifetime of $^{182}$Hf, these different components would have variable $^{182}$W values too (Kruijer and Kleine 2017). Instead, various lunar lithologies have the same $\varepsilon^{182}$W excess of +0.2 to +0.3 (Fig. 2D), implying that silicate differentiation of the Moon took place after significant $^{182}$Hf decay (i.e., later than 60-70 Myr after solar system formation; e.g., Touboul et al. 2007, 2009; Kruijer and Kleine 2017).

Touboul et al. (2015) and Kruijer et al. (2015) argued that the currently observed difference in $\varepsilon^{182}$W between lunar and terrestrial rocks reflects disproportional late accretion of meteoritic material by the Earth and Moon. The abundances of highly siderophile elements (HSE) in the lunar and terrestrial mantles suggest that these were delivered by an approximately chondritic late accretion after core formation had ceased (e.g., Day et al. 2007). Proportionally, the implied fraction delivered to the Earth during late accretion (~0.5% by mass; e.g., Bottke et al. 2010) is much larger than for the Moon. Chondrites have an $\varepsilon^{182}$W value near -1.9, meaning that late addition of such a component to the Earth and Moon would have lowered their mantle $\varepsilon^{182}$W values, but the effect is much larger for the Earth due to its greater inferred late accreted mass. Isotopic mass balance indicates that the current observed Earth-Moon difference of ~0.2 to 0.4 in $\varepsilon^{182}$W could be completely accounted for by this process (e.g., Kruijer et al. 2015; Touboul et al. 2015). It has thus been argued that the most "parsimonious" (i.e., probable) interpretation is that the terrestrial and lunar mantles initially had equal W isotopic compositions, with their current modest difference due to chondritic late accretion (e.g., Touboul et al. 2015; Kruijer et al. 2015; Kruijer and Kleine 2017). Alternatively, the small Earth-Moon W isotopic difference may have been generated by lunar core formation while $^{182}$Hf was extant. If Moon's origin and lunar core formation occurred within the lifetime of $^{182}$Hf (<55 Ma), radiogenic $^{182}$W would subsequently form in the silicate Moon, leaving it slightly more radiogenic than Earth, as observed (Thiemens et al. 2019).

The inferred initially equal Earth-Moon W compositions has become an important constraint on lunar origin that is perhaps even more challenging to explain than Earth-Moon isotopic similarities in other elements. Even a giant impactor that was Earth-like in $\Delta^{17}$O would still generally be expected to have a different W isotopic composition established by the details and timing of its own core formation. Dauphas et al. (2014a) calculated the required $\varepsilon^{182}$W and Hf/W values of the protoEarth and impactor mantle required to produce the similar Hf/W and $\varepsilon^{182}$W values of the lunar and terrestrial mantles in the context of the canonical model without equilibration. The conclusion of this analysis is that fine tuning is needed to match observations (Fig. 6 of Dauphas et al. 2014a). If protolunar material was derived overwhelmingly from the proto-Earth's mantle, a match in the W isotopic compositions of the silicate portions of the protolunar material and the Earth's mantle could result. However, incorporation of a very small fraction of material from Theia's core into the Moon could cause too large of a divergence in Earth-Moon W isotopic compositions even in this case (Kruijer and Kleine 2017; §4.2.2). As a result, planet-disk mixing may be needed to equilibrate W compositions after the Moon-forming impact but before the Moon accreted (e.g., Pahlevan and Stevenson 2007; Pahlevan 2018; Lock et al. 2018). We return to this important issue in §5.1 to §5.2.

### *2.5.2 Timing of lunar formation and LMO differentiation.*

Geochronological data on lunar samples are not straightforward to interpret and much controversy remains regarding the timing of lunar formation and LMO differentiation. As discussed in Sect. 2.5.1., depending on how $\varepsilon^{182}$W data are interpreted, they can indicate that the Moon formed by a giant impact approximately 50 Myr after solar system birth (Thiemens et al. 2019) or they can convey no geochronological information (Kruijer et al. 2015; Touboul et al. 2015). A significant source of uncertainty in establishing the age of the Moon and the completion of core formation on Earth using $^{182}$Hf-$^{182}$W systematics is the uncertainty on the extent of metal-silicate equilibration during impacts. For example, Fischer and Nimmo (2018) showed that existing data could be reproduced with a Moon-forming impact between 10-175 Myr after solar system birth (SSB), but most early impact scenarios are associated with a late veneer mass that exceeds ~0.5% of Earth's mass, so a range of ~60-175 Myr appears more likely (Jacobson et al. 2014; Fischer and Nimmo 2018).

The time of LMO crystallization is also uncertain and available data not straightforward to interpret. Different lunar lithologies have identical $^{182}$W isotopic compositions, which would indicate that LMO differentiation occurred after decay of $^{182}$Hf; more than 70 Myr after T$_o$. Supporting this view are a wealth of new data on $^{146}$Sm-$^{142}$Nd that yield a model age of global lunar differentiation of ~190 Myr after T$_o$ (Carlson 2019; Borg et al. 2019 and references therein). Other data however contradict such a late timeframe for global magmatism on the Moon.

Radiometric dating of ferroan anorthosites (FANs) using the Sm-Nd system resulted in a combined older age of 111 ±40 Myr after $T_o$ (Norman et al. 2003). Barboni et al. (2017) obtained model ages of urKREEP from $^{176}$Lu-$^{176}$Hf systematics of zircons that are as old as ~60 Myr after $T_o$. The time span between Moon formation and lunar crust formation could have been up to ~ 200 Myr due to tidal heating from Earth (Elkins-Tanton, et al. 2011b), so the age of the Moon's crust cannot help pin down the time of the giant impact beyond providing a lower limit.

To summarize, there is considerable uncertainty on the timing of Moon formed, with a very conservative range of 10-190 Myr, and a more plausible range of 50-150 Myr (Fischer and Nimmo 2018; Jacobson et al. 2014; Thiemens et al. 2019; see also Gaffney et al., 202X, this volume).

### 2.6. Summary

Since *New Views of the Moon* (2006), Earth and Moon bulk and isotopic compositions are known to greater resolution. Some volatile element concentrations in the lunar mantle may be similar to or higher than Earth's mantle, and higher than previously expected (H, F, Cl), while others (S,C) are lower in the BSM than in the BSE. Volatile metals have generally comparable depletions to volatile lithophile elements, but are more severely depleted for metals of lowest condensation temperatures (e.g., Pb, Zn, In, Cd). The generally heavy isotopic composition of volatile elements such as Ga, K, Rb, Cu, and perhaps Zn may help to constrain the origin of the volatile element depletions. Many elements have identical isotopic compositions in the Earth and Moon or only very small differences (Si, Ti, Ca, O, and W). However, the bulk FeO content of the lunar mantle appears to be 1.1 to 1.4 times that of the terrestrial mantle – somewhat higher than expected from an Earth-Moon compositional match.

New W measurements have been interpreted to imply that the Earth and Moon had equal W isotopic compositions when the Moon formed, providing a strong new constraint on lunar origin models. The small $^{182}$W excesses in the lunar mantle would then be explained either by core formation in a slightly older Moon (if lunar core formation occurred 40-60 Ma after $T_0$), or by disproportionate late accretion onto the post-impact Moon and Earth (if the Moon formed $\geq$ 60-70 Ma after $T_0$). The former scenario is supported by the 4.51 Ga zircon ages derived by Barboni et al. (2017), and the oldest Sm-Nd ages from 62255. The latter scenario is supported by evidence for a 'young Moon' from Sm–Nd model ages, which indicate lunar crust formation ages that vary from 4.35 to 4.45 Ga (Borg et al. 2015, 2019). The time of lunar magma ocean crystallization is uncertain, with some evidence suggesting that it took place shortly after the giant impact and other measurements supporting global differentiation ~190 Myr after solar system birth.

## 3. GIANT IMPACT SCENARIOS

Impact origin studies share a common goal: to identify collisional scenarios that can account for the properties of the Earth-Moon system, and among such scenarios, to identify those that appear most plausible based on our understanding of Earth's formation. Accomplishing this has proved challenging and has inspired a variety of impact models.

### 3.1 General constraints and methods

***3.1.1 Constraints.*** The current Earth-Moon system angular momentum is $L_{EM} = 3.5 \times 10^{41}$ g cm$^2$ s$^{-1}$. Tidal interaction between the Earth and Moon has caused the Earth's spin to slow and the Moon's orbit to expand while conserving angular momentum. A newly-formed Moon orbiting just outside the Roche limit implies an initial terrestrial day of about 5 hours if the primordial system angular momentum was comparable to $L_{EM}$.

An oblique collision(s) is a natural source of Earth's early rapid spin. The angular momentum delivered by an impactor of mass $M_i = \gamma M_T$, where $M_T$ is the total colliding mass, is (e.g., Canup 2004b)

$$L_i = b M_T^{5/3} f(\gamma) \sqrt{\frac{2G}{(4\pi\rho/3)^{1/3}}} \left(\frac{v_i}{v_{esc}}\right) \approx 1.3 L_{EM} b \left(\frac{M_T}{M_\oplus}\right)^{5/3} \left(\frac{f(\gamma)}{0.1}\right) \left(\frac{v_i}{v_{esc}}\right) \quad (3.1)$$

where $b = \sin \xi$ is a scaled impact parameter, $\xi$ is the impact angle (with $\xi = 0°$ a head-on impact), $v_i$ and $v_{esc}$ are the impact and mutual escape velocities, respectively, $\rho$ is the average density of the impactor and target (assumed in this expression to be equal), and $f(\gamma) = \gamma(1-\gamma)\sqrt{\gamma^{1/3}(1-\gamma)^{1/3}}$, where $\gamma \equiv M_i/M_T$. For $\gamma \leq 0.25$, $f(\gamma) \sim \gamma$. It was previously thought that the Earth-Moon system immediately after a Moon-forming impact would need to have an angular momentum close to $L_{EM}$ because subsequent angular momentum alteration due to, e.g., mass escape during lunar accretion, direct solar tides on the Earth, effects of smaller later impacts, and/or short-term capture into the evection resonance was expected to be minor, on the order of a few to 10%. However, dynamical mechanisms have been identified that could remove substantial angular momentum from the Earth-Moon system after the Moon formed, so that the initial system angular momentum could potentially have been as high as $\approx$ 2 to 3$L_{EM}$ (Ćuk & Stewart 2012; Wisdom and Tian 2015; Ćuk et al. 2016; Tian et al. 2017; Rufu and Canup 2020; see §4.7).

A successful single impact scenario must produce a disk with enough mass and angular momentum to later accrete into a lunar-mass Moon, with mass $M_L = 0.0123 M_\oplus$, orbiting outside the Roche limit ($a_R = 2.9 R_\oplus$ for lunar-density materials). From conservation of mass and angular momentum, the mass of a moon, $M_M$, that can accrete at a distance $a_M = X_a a_R$ from a disk of mass $M_D$ and angular momentum $L_D$ is

$$\frac{M_M}{M_D} \approx C_1 \left(\frac{L_D}{M_D \sqrt{G M_\oplus a_R}}\right) - C_2 - C_3 \left(\frac{M_{esc}}{M_D}\right) \quad (3.2)$$

where $M_{esc}$ is the mass that escapes as the moon accretes (Ida et al. 1997), and coefficients $C_1$, $C_2$ and $C_3$ are determined by $X_a$ and the mean specific angular momenta of material that escapes or collides with Earth. Equation (3.2) provides a simple, approximate way to determine whether a disc produced by a given impact could later yield a lunar mass Moon. It is physically invalid for cases that would give $M_M/M_D > 1$; these correspond to initial disks with specific angular momenta too high to be consistent with the assumption that the final moon forms at $X_a a_R$. Early pure N-body lunar accretion simulations (Ida et al. 1997; Kokubo et al. 2000) found $\langle a_M \rangle = 1.3 a_R$, while more recent hybrid N-body simulations find $\langle a_M \rangle = 2.15 a_R$ (Salmon and Canup 2012), implying less efficient accretion (§4.5). An initial disk with between about 1.3 and 3 lunar masses is needed to yield a lunar-mass Moon across typical post-impact conditions.

The mantle of the Moon is relatively iron-rich but overall, the Moon contains less than ~ 10% of its mass in the element iron due to its small core. It is typically assumed that a successful impact must produce an iron-depleted disk, although alternatively the disk could evolve into an iron-depleted state via post-impact vapor mixing with the Earth's mantle (§4.2).

As discussed in Section 2, the Moon and the Earth's mantle share essentially identical isotopic compositions across many elements, while meteorites from Mars and the asteroid belt differ substantially. For a giant impactor with an isotopic composition different from that of the Earth, but with similar bulk elemental concentrations, the compositional difference between the silicate components of the resulting disk and the post-impact Earth can be quantified as

$$\delta f_T (\%) \equiv \left[\frac{F_{D,tar}}{F_{P,tar}} - 1\right] \times 100 \quad (3.3)$$

where $F_{D,tar}$ and $F_{P,tar}$ are the mass fractions of the silicate portions of the disk and the post-impact planet that were derived from the pre-impact target's mantle, respectively. Identical disk-planet compositions would have $\delta f_T = 0$, while those with disks that contain proportionally more impactor [target] material would have $\delta f_T < 0$ [$\delta f_T > 0$]. In the absence of post-impact equilibration (§4.2), the disk's initial composition would be expressed in that of the Moon. In this case, explaining the Earth-Moon oxygen similarity (i.e., $|\Delta^{17}O| < 10$ ppm) requires $|\delta f_T| < 5\%$ for an impactor that was Mars-like in its isotopic composition (e.g., Canup 2012). The case has been made that an isotopically Mars-like impactor is most likely (Pahlevan & Stevenson 2007), although counter arguments for an Earth-like composition impactor have also been advanced (e.g., Belbruno and Gott 2005; Dauphas et al. 2014a,b; Mastrobuono-Battisti et al. 2015; Kortenkamp and Hartmann 2016; Dauphas 2017; see §5.1). Explaining the Earth-Moon similarity in W imposes a separate and stringent constraint (see §4.2.2 and §5.1).

*3.1.2 Methods.* Ejecta on circum-planetary orbits can result from vaporization-related pressure gradients and/or gravitational torques due to interactions among the ejected material or between the ejecta and the distorted shape of the post-impact planet. Modeling the production of disks by impacts thus requires a 3D hydrodynamical approach that includes phase changes and explicit self-gravity. The most commonly utilized method is smoothed particle hydrodynamics (SPH), which describes the colliding objects by a multitude of particles that are evolved due to self-gravity, pressure forces, and shock dissipation (e.g., Benz et al. 1989). Each particle is assigned a silicate or iron composition to represent the mantle or core, and a corresponding equation of state. A limited number of lunar forming impact simulations have also been performed using grid-base Eulerian codes (Wada et al. 2006; Canup et al. 2013).

## 3.2 Canonical impact

A roughly Mars-mass impactor that collides with Earth at a low velocity (comparable to $v_{esc}$) can produce an iron-depleted Moon and a planet-disk system whose angular momentum is ≈ $L_{EM}$ (Canup and Asphaug 2001; Canup 2004a,b, 2008, 2014). The required impactor mass and velocity are common in models of terrestrial planet accretion, and the needed collision angle (≈ 40 to 50°; Canup 2004, 2008) is centered on the most probable angle for randomly oriented impacts (45°), so that about 20% of impacts fall within the needed range.

The key difficulty for a canonical impact is accounting for Earth-Moon isotopic similarities. In canonical impacts, most of the disk, typically ∼ 70 to 80%, originates from the impactor's mantle, producing $\delta f_T$ ∼ −90 to − 70% (e.g., Fig. 1 in Canup 2014). This is a vastly larger disk-planet compositional difference than the $|\delta f_T| < 5\%$ required to explain the Earth-Moon similarity in oxygen for an isotopically Mars-like impactor in O. This overall result has not been strongly affected by orders-of-magnitude increases in numerical resolution, and appears in simulations using both standard SPH and the grid code CTH (e.g., Canup et al. 2013). Pre-impact rotation in the colliding bodies modestly affects the portion of the disk derived from the impactor, with this percentage varying from 60 to 90% (and $\delta f_T \leq -50\%$) for impacts involving pre-impact rotation if the final system angular momentum is constrained to be near $L_{EM}$ (Canup 2008).

However, Hosono et al. (2019) found that a canonical impact can produce a disk formed primarily from the Earth's mantle if 1) Earth had a surface magma ocean at the time of the impact, and 2) a modified SPH code is utilized. They argue that a complex silicate melt is best described by a hard-sphere equation of state (EOS) to account for increased heating upon shock compression. With a hard-sphere EOS to describe the Earth's magma ocean and the new SPH code, their simulations of canonical impacts produced disks that were both massive ($M_D \geq M_M$) and Earth-dominated, having on average only ∼ 30% impactor material by mass ($\delta f_T \approx -17\%$). In contrast, their simulations with standard SPH and a hard-sphere EOS produced impactor-dominated disks (see §5.4), similar to earlier results.

In general, a canonical Moon-forming impact requires either an impactor with an Earth-like isotopic composition, or that the compositional signature of the impactor in the disk was erased by vapor mixing between the post-impact disk and planet, a process known as equilibration (Pahlevan and Stevenson 2007; §4.2). Given current understanding of terrestrial planet accretion, the first solution appears improbable (Pahlevan and Stevenson 2007; §5.1), while the likelihood of the second is highly uncertain (§5.2).

### 3.3 Hit-and-run impact

Less oblique, higher-velocity impacts can create disks with proportionally more target material. For higher $v_i$, there can also be substantial escaping angular momentum, allowing for larger values of $L_i$ and therefore larger impactors. Reufer et al. (2012) identified a regime of hit-and-run collisions with impact angles of 30 to 40°, $1.2 \leq (v_i/v_{esc}) \leq 1.4$, and more massive impactors ($\gamma \approx 0.2$) that produces disks with 40-60% of their mass derived from the impactor, leading to $\delta f_T$ ∼ −50 to − 35% (e.g., Fig. 1 in Canup 2014). This is a smaller planet-disk compositional difference than in canonical impacts, but is still larger than needed to explain Earth-Moon isotopic similarities for a Mars-like impactor. The most successful hit-and-run impacts that produce disks massive enough to produce a lunar-mass Moon also yield systems with angular momenta ∼ 1.3 to 1.4 $L_{EM}$ (Reufer et al. 2012; Canup 2014; Deng et al. 2019), so that a mechanism to extract angular momenta after the impact is required (§4.7).

### 3.4 Fast-spinning Earth impact

With the goal of providing a more natural explanation for the Earth-Moon isotopic similarities, recent works have considered impacts that produce much higher angular momenta systems (Ćuk and Stewart 2012; Canup 2012; Lock et al. 2018). Among these, two impact configurations have been identified that can produce a post-impact planet and disk with similar proportions of their silicate material derived from the two colliding bodies (i.e., with $|\delta f_T| \sim 1$ to 10%), which could satisfy the oxygen isotopic data even in the limit of a Mars-like composition impactor. The first considers an impact into a proto-Earth whose initial spin angular momentum greatly exceeds $L_{EM}$ (discussed here), and the second considers a collision between two half-Earth mass bodies (§3.5). Both require a subsequent dynamical mechanism to bring the system to the present-day angular momentum (§4.7).

*N*-body simulations of terrestrial planet accretion predict that Earth-like planets form with rapid rotation due to the effects of giant impacts if fragmentation is neglected, an assumption that would tend to provide an upper limit on spin rate (Kokubo and Genda 2010). Ćuk and Stewart (2012) examined impacts into rapidly rotating, approximately Earth-mass targets whose rotation contained between about 2.0 and 3.0$L_{EM}$ before the Moon-forming impact, with the upper end of this range being approximately the spin stability limit. Relatively small, fast impactors ($M_i = 0.026$ to $0.1 M_\oplus$ and $v_i = 1.5$ to $3 v_{esc}$) with impact angles between 0 and −20° (i.e., that were retrograde with respect to the planet's pre-impact rotation) produced potential Moon-forming disks with $|\delta f_T|$ between 4 and 15%, with the disk material originating overwhelmingly from the target's mantle.

A fast-spinning Earth impact, as well as other high angular momentum/high energy impacts (§3.5-§3.6), has a specific impact energy about an order of magnitude larger than a canonical impact. Most of the disk-like regions are pure vapor, as opposed to a disk produced by a canonical impact that is initially dominated by melt, leading to differences in post-impact structure and evolution (§4.4).

## 3.5 Half-Earth impact

If the mass of the Moon-forming impactor is $\ll M_\oplus$, the fraction of the post-impact planet derived from the target is near unity ($F_{P,tar} \sim 1$). Forming a disk with a very small compositional difference from the planet then requires $F_{D,tar} \sim 1$ from eqn. (3.3), *i.e.*, a disk derived overwhelmingly from the target. However, the more general requirement to produce a disk and planet with equal silicate compositions is to have $(F_{P,tar}/F_{D,tar}) \sim 1$. The limiting case of a completely symmetric collision by two objects, each having one-half the Earth's mass, would produce a disk and final planet that both contain 50% target and 50% impactor material. For this case, $F_{P,tar} = F_{D,tar} = 0.5$, $(F_{P,tar}/F_{D,tar}) = 1$, and $\delta f_T = 0$, so that the disk and planet would have equal silicate compositions even though the disk was not derived primarily from the protoearth.

Somewhat asymmetric collisions with an impactor whose mass is > 40% of the Earth's mass can produce disk-planet compositional differences as small as $|\delta f_T| \sim 1$ to 10% (Canup 2012). Successful cases occur across a range of impact speeds ($1 \leq v_i/v_{esc} \leq 1.6$) and impact parameters ($0.35 \leq b \leq 0.7$), and do not require a particular pre-impact rotational state for the protoearth. A similar type of collision between comparably sized bodies is thought to be responsible for the Pluto-Charon binary (Canup 2005). A half-Earth impact can also create a more massive disk than other scenarios, with $M_D \sim 2$ to $5 M_L$, which is advantageous for forming a sufficiently massive moon if accretion is inefficient (e.g., Lock et al. 2018, their Fig. 7). It produces an Earth-disk system whose angular momentum is ≈ 1.8 to $2.7L_{EM}$, a large excess that must be subsequently removed.

## 3.6 General high-angular momentum/high-energy impact

The fast-spinning Earth and half-Earth scenarios are limiting cases of high angular momentum impacts that can directly produce disk-planet systems with nearly equal silicate compositions. However, a broader range of high angular momentum impacts would be viable if equilibration removed moderate disk-planet compositional differences.

Lock and Stewart (2017) and Lock et al. (2018) found that a wide range of giant impacts produce a so-called synestia – a partially vaporized, rapidly rotating and distended planetary body whose internal energy and angular momentum exceeds that of a corotating object (see §4.4). Equilibration within a synestia may be achieved by multiphase convective/turbulent mixing, particularly within its high entropy regions (Lock et al. 2018). Equilibration could allow high-angular momentum impacts that produce massive, impactor-dominated disks to ultimately yield moons with Earth-like compositions.

A substantial fraction of impacts that produce post-impact bodies with angular momenta >$1.7L_{EM}$ (including fast-spinning Earth and half-Earth impacts) are likely to form synestias, but impacts that produce lower angular momentum systems might also produce synestias if they are very high-energy (see §4.4). Most Earth-like planets probably experienced one or more synestia-producing impacts during their late accretion (Quintana et al. 2016; Lock and Stewart 2017).

## 3.7 Multiple impacts

Simulations suggest that Earth experienced multiple planetary-scale impacts during its final accretion (e.g., Agnor et al. 1999). That the Moon could be the cumulative result of multiple collisions was first proposed by Ringwood (1989), but has only recently been explored with modern methods. In the Rufu et al. (2017) multiple impact model, the Moon forms from a sequence of medium- to large-size collisions by impactors of mass $10^{-2}$ to $10^{-1} M_\oplus$. Each impact generates a sub-lunar mass satellite that migrates outward due to tidal interaction, faster at first, and slower as the body retreats away from the proto-Earth. A subsequent impact produces another inner moonlet, whose tidal expansion can cause it to merge with the prior outer moon. In this way the Moon can be built up by moonlets produced by a series of impacts. Although the impacts occur with random orientation, about 10% of histories produce a final Earth that is rotating appropriately rapidly (Rufu et al. 2017; see §5.4).

In comparison with a canonical impact, the Rufu et al. model considers smaller impactors, and a larger range of impact angles (from head-on to near-grazing) and velocities (1 to $4v_{esc}$). Nearly head-on and/or high velocity impactors produce disks with the highest proportion of target material, and protolunar material derived from such events can increase the likelihood of a compositional match between the Earth and its final Moon. In addition, as the number of impactors and associated moonlets increases, the variance of the difference between the compositions of the final Earth and Moon tends to decrease as $\sqrt{N}$, with the compositions of both the final Earth and Moon approaching that of the mean planetesimal neighborhood. Using Monte Carlo simulations involving expected populations of sub-Mars sized impactors (*e.g.* Raymond et al. 2009), together with results from SPH simulations and the assumption that all moonlets formed by subsequent impacts merge, Rufu et al. (2017) find that a sequence of $N \sim 20$ to 30 impacts can form a lunar-sized satellite, with substantial percentages of their impact histories producing an isotopically similar Earth-Moon pair.

## 4. POST-IMPACT EVOLUTION

Here we discuss recent investigations into the post-impact evolution of the planet and its orbiting material. This evolution is complex, requiring a multi-disciplinary description, and must ultimately produce an Earth-Moon system consistent with a variety of dynamical and compositional constraints.

### 4.1. Mixing during a Moon-forming impact and Earth's mantle heterogeneity

A key geochemical system relevant to the early Earth is Hf-W. By convention, Earth's current mantle has $\varepsilon^{182}W = 0$, while chondrites have $\varepsilon^{182}W = -1.9$ (e.g., Schoenberg et al. 2002; Kleine et al. 2002; Yin et al. 2002). However, recent studies find that some terrestrial Archean samples have positive W anomalies relative to the BSE ($\mu^{182}W = 20$ ppm, Willbold et al. 2011; Touboul et al. 2012), some modern flood basalts have even higher values ($\mu^{182}W \approx 50$ ppm, Rizo et al. 2016), and others have negative values (Mundl et al. 2017). Such ancient anomalies have been preserved until today and may be related to seismic anomalies observed at the core-mantle boundary (large low-shear velocity provinces, LLSVPs, e.g., Garnero and McNamara 2008).

In addition to tungsten, other systems may also indicate the preservation of early terrestrial mantle heterogeneity. Mukhopadhyay (2012) suggests that differences in $^{20}Ne/^{22}Ne$ and $^{129}Xe/^{130}Xe$ (where $^{22}Ne$, $^{20}Ne$, and $^{130}Xe$ are stable and $^{129}Xe$ is the daughter product of $^{129}I$ with $t_{1/2} = 16$ Myr) between OIB (Ocean Island Basalt) and MORB (Mid-Ocean Ridge Basalt) can be explained by separation of their reservoirs by 4.45 Ga. The observation that $^{20}Ne/^{22}Ne$ ratios in fluid inclusions in plume-related rocks are similar to solar values (Yokochi and Marty 2004) suggests that the growing Earth's mantle may have captured nebular gas while the solar nebula was present (e.g., Ikoma and Genda 2006) and that the mantle has never since been completely homogenized.

The Moon-forming impact is generally thought to have occurred ~ 60-70 Myr after the formation of the earliest solar system solids (§2.5). However, given uncertainties in metal-silicate equilibration during Earth's accretion, available $\varepsilon^{182}W$ data could be explained with a formation time as broad as 10-175 Myr (Fischer and Nimmo 2018). If the Moon formed after the observed mantle heterogeneities developed, their preservation implies that the Moon forming event(s) did not homogenize the Earth's mantle. A canonical impact and the multiple impact model appear consistent with a non-fully mixed mantle. The much higher energy fast-spinning Earth impact may also preserve some mantle heterogeneities in the cold region opposite to the impactor point (Figure 3b), while a half-Earth impact appears likely to dynamically mix the mantle (Nakajima and Stevenson 2015; Fig. 3c) and erase mantle heterogeneities.

[Figure 3]

More work is needed to assess this complex issue. It remains uncertain how much mixing of Earth's mantle is allowable, given that the total mass represented by the above described isotopic heterogeneities may be tiny. If the deep mantle was compositionally stratified, this portion may have survived even an energetic Moon-forming impact (e.g., Nakajima and Stevenson 2015; Lock et al. 2018; C. R. M. Jackson et al. 2018). Moreover, it is possible that at least some of the observed mantle heterogeneities could have developed after Moon formation. Potential explanations for the tungsten anomaly include: (1) materials added after the Moon-forming impact (so called late veneer) (e.g., Willbold et al. 2011; Marchi et al. 2018), (2) core-mantle interaction (e.g., Mundl et al. 2017), and (3) incomplete mixing of the proto-Earth and impactor materials (Rizo et al. 2016). Nevertheless, the establishment of primordial Ne isotopic signatures would seem to predate the removal of the solar nebula, requiring incomplete homogenization of the Earth's mantle by all post-nebular giant impacts. More work is needed to investigate the mantle's time evolution to quantify the likelihood of preservation of mantle heterogeneities over geological time scales (e.g., van Keken et al. 2002, Tackley 2012, Debaille et al. 2013; Petõ et al. 2013; Caracausi et al. 2016; Foley and Rizo 2017).

**4.2 Disk-planet equilibration**

Proposed Moon-forming giant impacts release enormous gravitational potential energy, typically 4 to $10 \times 10^{31}$ J, an energy budget that cannot be accommodated without melting and partly vaporizing a significant fraction of the post-impact system. Both the fluid state of the post-impact Earth-disk system and the prolonged epoch required for its cooling (~$10^2$ to $10^3$ years) has led to the proposal that the silicate Earth and proto-lunar material were essentially homogenized into a single isotopic reservoir through fluid-dynamical convective mixing, a process referred to as equilibration (Pahlevan and Stevenson 2007). Equilibration would allow pre-lunar material to acquire the Earth's isotopic composition before the Moon was assembled, offering an attractive explanation for Earth-Moon isotopic similarities. At a minimum, equilibration requires vigorous radial mixing in the proto-lunar disk at a rate that is faster than angular momentum transport (Pahlevan and Stevenson 2007; see §5.2) and vertical convective mixing of a significant fraction of the post-impact Earth (Nakajima and Stevenson 2015; Lock et al. 2018; see §5.2).

*4.2.1. Theoretical developments.* Recent developments relevant to the potential for Earth-disk equilibration include a number of scenarios ranging from those with full isotopic equilibration to more heterogeneous mixtures (equilibrated and unequilibrated reservoirs), and the description of the conditions and the observable signatures of system-wide mixing.

First, are scenarios in which the post-impact silicate Earth experiences convection from deep layers to low-pressure atmospheric regions connected to the proto-lunar disk. This leads to homogenization of all chemical and isotopic tracers in the silicate Earth and Moon. This equilibration scenario could arise if convective motions are not inhibited by a magma-ocean-atmosphere interface. Liquid at high pressure is carried to a two-phase region where bubbles begin to appear, with droplets then carried by advection into a partly-vaporized atmosphere-like region, allowing even elements that do not significantly vaporize to equilibrate via suspended droplet exchange (Pahlevan et al. 2011). Alternatively, turbulent mixing may have occurred at earlier times in the thermal history when the post-impact Earth existed in the form of a supercritical fluid, i.e. there was a single fluid phase and therefore no possibility for phase separation (§3.6 and §4.4; Lock et al. 2018). However, the ability of such a scenario to dilute isotopic differences may be limited because the outer layers of the post-impact Earth typically have the highest entropy and may not mix readily with deeper, lower entropy layers (see Nakajima & Stevenson 2015) (§5.2).

Second, is the proposal that the deep Moon may have accreted from the material initially injected into the outer disk and not subject to equilibration, overlain by Earth-equilibrated inner disk material represented by the sampled Moon (Salmon and Canup 2012; Canup et al. 2015; §4.6). Third, are constraints from SPH simulations on the initial thermal state of the post-impact Earth that limit the degree to which the deep Earth could have participated in the Earth-disk equilibration process (Nakajima and Stevenson 2015; Lock et al. 2018). Fourth, is the proposal that the velocity shear between the post-impact Earth and innermost disk may have facilitated mixing at the Earth-disk interface (Gammie et al. 2016). And fifth, is a proposal that liquid-vapor equilibration can fractionate isotopes even in the high-T post-impact environment, leading to coupled chemical and isotopic silicate Earth-Moon differences (Pahlevan et al. 2011).

*4.2.2. Empirical constraints.* Earth-Moon isotopic homogeneities, initially emphasized for oxygen (e.g., Wiechert et al. 2001; Pahlevan and Stevenson 2007) are now known to extend to (§2.3): (1) refractory elements like titanium (Zhang et al. 2012) and perhaps calcium (Schiller et al. 2018), requiring either partial vaporization and equilibration of even refractory elements (Zhang et al. 2012), droplet-droplet exchange in vapor suspensions (Pahlevan et al. 2011), and/or an Earth-like impactor in refractory elements (Dauphas et al. 2014a; Dauphas 2017); (2) siderophile elements like tungsten (Kruijer et al. 2015; Touboul et al. 2015), whose inferred similarity in the post-giant impact Earth and Moon may require Earth-disk equilibration after metal-silicate equilibration in the terrestrial magma ocean and the proto-lunar disk (see below), and (3) major elements such as silicon that display distinct bonding environments in the liquid and vapor, which may therefore fractionate following liquid-vapor separation. Per (3), determination of the bulk lunar composition via other methods can then furnish tests of the equilibration scenario (Pahlevan et al. 2011; Armytage et al. 2012; Sakai et al. 2014; see §5.2).

The isotopic homogeneity of tungsten is of particular interest (e.g., Pahlevan 2014). Consider the limiting case of a disk whose silicate component had the same W isotopic composition as the BSE, e.g., a disk formed overwhelmingly from the Earth's mantle. If the disk also contained even a percent or so of its mass in metal originating from the impactor's core, as is consistent with impact simulation results, the expected negative $\varepsilon^{182}W$ composition of this core material, together with its high absolute abundance of siderophile W, would likely generate a divergence in the Earth-disk tungsten isotopic compositions larger than that inferred in recent work (e.g., Kruijer and Kleine 2017, their Fig. 6a). Thus even high-angular-momentum scenarios that produce disks with $|\delta f_T| \sim 1$ to 10% (Ćuk and Stewart 2012; Canup 2012) would require either a fortuitous impactor and protoearth compositional relationship in W (e.g., Dauphas et al. 2014a) or disk-planet equilibration to reproduce the inferred similar W compositions of the initial Earth and Moon (see §5).

## 4.3. Proto-lunar disk evolution

The evolution of the proto-lunar disk is arguably less well-understood than the mechanics of the impact itself. This has motivated a number of recent efforts to better understand this evolution and make new connections with cosmochemical observables.

*4.3.1. Initial thermal states*. "Canonical" impacts produce disks that are moderately (~10 to 20%) vaporized, while high-angular-momentum scenarios generate highly (80-90%) vaporized disks (e.g., Fig. 4 of Nakajima and Stevenson 2014; Lock et al. 2018; see also §4.4). This clear contrast might be expected to yield distinct predictions that permit discrimination between the proposed impact scenarios. However, such one-to-one mapping of initial thermal states to lunar observables is hampered by an incomplete understanding of the intervening viscous evolution that could dramatically affect both the radial distribution of a disk's mass and its thermal state.

*4.3.2. Thermal and viscous evolution*. As with all astrophysical disks, the proto-lunar disk may be subject to viscous evolution, causing radial spreading and self-heating of the disk via gravitational energy release, given a mechanism that translates shear motions into a viscous couple (Lynden-Bell and Pringle 1974). For a massive disk of condensed material interior to the Roche limit, self-gravity leads to the formation of temporary clumps that are continually disrupted by the planet's gravity, producing a large viscosity (Ward and Cameron 1978; Daisaka and Ida 1999; Takeda and Ida 2001; Daisaka et al. 2001). Recent developments include the concept that the proto-lunar melt largely settled to the disk mid-plane, resulting within the Roche limit in vigorous gravitational instability near the mid-plane that produced a viscosity and associated heating, with the melt layer surrounded by a gravitationally stable silicate vapor atmosphere (Ward 2012, 2014, 2017). This model is the stratified end-member to the vertically well-mixed, single-column isentropic structures considered in Thompson and Stevenson (1988) and Pahlevan et al. (2011). It has been argued that a stratified disk structure is most probable (Machida and Abe 2004; Ward 2012, 2014, 2017).

The one-dimensional numerical model of a stratified disk's thermal evolution by Charnoz and Michaut (2015) demonstrated that viscous heating and radiative cooling never balance instantaneously as assumed by Thompson and Stevenson (1988). Ward (2014, 2017) argued that this difference between energy loss via radiative cooling and that supplied by viscous heating will result in silicate condensation/vaporization until the Roche-interior melt layer achieves a quasi-steady state total mass of about $0.14M_L$ (Ward 2014). This is the melt layer mass for which the production of heat via viscous dissipation due to gravitational instability in the melt is balanced by radiative cooling from the surfaces of the silicate vapor atmosphere (Ward 2014). If the inner melt layer mass is initially larger than this value, viscous heating will drive additional vaporization until quasi-steady state melt layer mass is achieved. The implication is that, for a nominal lunar-mass inner disk, the inner disk vapor mass fraction may rapidly increase to up to >80% even if the disk initially contains much less vapor. More complex evolutions result if the innermost disk is treated as an inviscid, incompressible fluid not subject to gravitational instability (Charnoz and Michaut 2015).

Global cooling of the proto-lunar disk requires that radiative cooling removes both the latent heat produced by vapor condensation and any gravitational energy viscously produced. If the outer disk does not experience substantial viscous heating, cooling there would occur on a shorter timescale than cooling in the inner disk due to its lower surface mass density (e.g., Canup et al. 2015; Lock et al. 2018).

However, if the proto-lunar disk is sufficiently hot to vaporize and partially ionize the alkali elements (e.g., Visscher and Fegley 2013), this may enable magnetic coupling in the vapor portions of the proto-lunar disk both inside and outside the Roche limit (Carballido et al. 2016; Gammie et al. 2016). Unlike a viscosity produced by gravitational instability, in which energy release makes the disk hotter and more gravitationally stable, magnetic coupling has the property of becoming more effective at hotter states. Hence, the possibility exists of a positive feedback in which the magnetically produced viscosity spreads and heats the disk, leading to more effective coupling and more rapid spreading in a runaway process (Gammie et al. 2016). Whether or not the protolunar disk vapor would have been magnetically coupled depends on the magnetic field strength (which is uncertain) and the disk's ionization fraction, where the latter is set by the vapor chemistry (e.g., Carballido et al. 2016) and the effect of melt droplets, which may decrease (e.g., Ward 2017) or increase (e.g., Desch and Turner 2015) ionization, depending on whether they are net absorbers or emitters of charge.

A related issue is the criterion by which the disk fragments into moonlets. Several different criteria have been adopted. First, it has been considered that the disk may have the same criterion for fragmentation as a particle disk, namely, the classical Roche radius (Ida et al. 1997; Salmon and Canup 2012; Lock et al. 2018). Alternatively, the two-phase (liquid-vapor) nature of the disk may define a critical radius – beyond the Roche radius – where the surface density declines to a critical value and gravitational instabilities grow on an orbital timescale, causing fragmentation (Thompson and Stevenson 1988). However, settling of droplets to the mid-plane may preclude such an evolution (e.g., Machida and Abe 2004; Ward 2012). Finally, in the case of a magnetically coupled disk, the disk will not cool and condense until after magnetic (and viscous) decoupling, which Gammie et al. (2016) argue would be delayed until the disk edge had expanded to ~$10R_\oplus$. The latter mass profile for the start of lunar accretion is different than considered in all prior models (e.g., Ida et al. 1997; Salmon and Canup 2012; Lock et al. 2018). We conclude that, so long as the state of stratification and mechanisms driving viscous evolution in the proto-lunar disk are uncertain, the criteria by which the disk fragments into moonlets should also be considered an open question.

*4.3.3. Open/closed system behavior*. A question relevant to disk evolution is whether the disk behaves as an open or closed system. Evaluation of the Jeans parameter for thermal escape reveals that the critical mean molecular weight is low, ~3 to 4 proton masses at 2000 K and $5R_\oplus$, or perhaps a factor of ~2 larger if the criterion for hydrodynamic escape is altered due to Keplerian rotation (Stevenson 1987, Desch and Taylor 2011). This implies that silicate vapor (with a molecular mass ≈ 20 amu) was gravitationally bound, and that even light hydrogen could undergo escape only if it post-dated silicate vapor condensation and also overwhelmed the abundance of other heavy volatile species (e.g., CO), and/or if the proto-lunar disk extended to much larger radii than generally considered (see §4.6). These requirements are stringent, and have led to the view that the proto-lunar disk is, in fact, closed with respect to thermal escape, or nearly so (Stevenson 1987; Canup et al. 2015; Pahlevan et al. 2016; Nakajima and Stevenson 2018). More recently, it has been proposed that the Earth-disk system experienced magnetized winds that significantly altered the system's angular momentum but not its mass budget (Gammie et al. 2016; §4.7). This issue requires further study.

**4.4 Evolution of a synestia**

During a giant impact, portions of the colliding bodies experience sufficiently high shock pressures to vaporize the silicate mantles upon release to pressures below the critical point (e.g., about 0.1 GPa; Kraus et al. 2012). Because the vapor density far below the critical point can be orders of magnitude smaller than that of condensed silicates, the formation of substantial amounts of rock vapor significantly expands the radius of a post-impact body. A giant impact also affects a planet's rotation. The shape of a rotating fluid body is difficult to determine except for the homogeneous density case (i.e., Maclaurin spheroids or Jacobi ellipsoids; Chandrasekhar 1969), and for bodies with strong gradients in density the structure must be calculated numerically (Lock and Stewart 2017). As Earth-like planets increase in angular momentum, their rotation rate increases and they become very oblate spheroids, with a ratio of equatorial to polar radii as large as two.

As the internal energy increases in a rotating planet and the radius expands, the angular velocity decreases to conserve angular momentum. When the angular velocity at the equator of the body intersects Keplerian orbital velocities, the planet has reached its *corotation limit* (CoRoL). Beyond this limit, a single, corotating body is not possible. Instead, bodies beyond the CoRoL typically comprise an inner, corotating region that is smoothly connected to an outer disk-like region, a type of object referred to as a *synestia* (Lock and Stewart 2017), so named for "syn-" or together and "Hestia", the Greek goddess of architecture and structures. The inner boundary of the disk-like region evolves with time and there can be significant mass and angular momentum transport both ways across the boundary.

[Figure 4]

The corotation limit depends on thermal state, angular momentum, total mass, and compositional layering, as shown in Figure 4 for thermally-stratified bodies such as those produced in giant impacts. For a given post-impact angular momentum, formation of a synestia requires that the specific entropy of the post-impact body exceeds a critical value; alternatively, for a fixed post-impact planet entropy, a synestia requires that a critical value of the post-impact angular momentum is exceeded. Most synestias are largely vapor or supercritical fluids around the continuous transition between the corotating and disk-like regions. An example planet with an angular momentum of 2 to 2.5$L_{EM}$ (comparable to that produced by a fast-spinning Earth or half-Earth impact) has a corotating region that extends to an equatorial radius of ~ 1.6$R_\oplus$, interior to the Roche limit (2.9$R_\oplus$) but substantially larger than Earth's current radius.

A principal difference between a synestia and the standard case of a relatively thin disk is the importance of thermal (gas) pressure. Radial force balance in a disk is set by hydrostatic and centrifugal equilibrium:

$$\frac{GM}{r^2} + \frac{1}{\rho}\frac{dp}{dr} - \omega^2 r = 0, \qquad (4.1)$$

where $G$ is the gravitational constant, $M$ is the mass of the planet/central structure, $r$ is cylindrical radius, $\rho$ and $p$ are the gas density and pressure, and $\omega$ is angular velocity. The ratio of the pressure gradient term to the gravitational term is of order $(H/r)^2$, where $H \sim c/\Omega$ is the disk's vertical scale height, $c$ is the mid-plane gas sound speed, and $\Omega$ is orbital frequency at distance $r$. Thus, for thick, hot disks (i.e., with $H/r \geq 0.3$), the pressure gradient term becomes appreciable (e.g., Nakajima and Stevenson 2014; Lock et al. 2018). A parcel of vapor at a given specific angular momentum will then achieve force balance at a substantially greater radius $r$ than would be calculated for a pure Keplerian orbit, with $r$ including pressure support being of order 30 to 40% larger than $r$ for a Keplerian orbit (Lock et al. 2018, their Fig. 2). As the disk-like portions of a synestia cooled and condensed, radial pressure-support would decrease and material would contract inward to a Keplerian orbit without changing its angular momentum. But prior to that time, radial motions during condensation and re-evaporation could drive radial mixing (Lock et al. 2018), a variation on the idea of planet-disk equilibration (Pahlevan and Stevenson 2007, see discussion in §4.2).

In addition, large condensates orbiting within a highly vaporized disk will experience a size-dependent drag due to a head-wind from the slower orbiting gas. A body of density $\rho$ and radius $R$ that is large enough to not be coupled to the gas will spiral inward due to gas drag on a timescale

$$\tau_{GD} \sim \frac{8}{3 C_D} \frac{1}{\Omega} \left(\frac{\rho R}{\sigma_G}\right) \left(\frac{r}{H}\right)^3 \tag{4.2}$$

where $C_D$ is a drag coefficient of order unity (*e.g.,* Canup and Ward 2002), $\Omega$ is orbital frequency, and $\sigma_G$ is the gas surface density. Gas drag is particularly fast for thick disks, due to the $(r/H)^3$ term; e.g., with $\sigma_G \sim 10^6 \text{g/cm}^2$ (equivalent to 0.5 lunar masses in gas spread evenly between 3 and 6 Earth radii) and $(H/r) \sim 0.3$, a 1-km condensate near the Roche limit would spiral inward in only $\sim 10$ hr.

[Figure 5]

Many types of high-energy/high angular momentum impacts generate synestias (Fig. 5), although canonical impacts do not. An impact-generated terrestrial synestia, which is differentiated into iron and silicate layers, is strongly thermally stratified with a low-entropy silicate layer, comprising perhaps 75% of the total silicate mass, below a high-entropy outer silicate layer with perhaps 25% of the silicate mass (Fig. 5). Mixing may rapidly homogenize compositions within the high-entropy outer regions, while the lower entropy regions might not effectively mix (see §5.2). Producing an isotopically similar Earth and Moon via mixing (Lock et al. 2018) would also require the mixing to extend out into the disk-like regions where the Moon forms, and that turbulent diffusion was more efficient than turbulent viscosity, a requirement of all planet-disk equilibration scenarios (Pahlevan and Stevenson 2007). Because the outer regions of a synestia are largely vaporized by the impact, a substantial gas pressure extends beyond the Roche limit. Thus, moonlets that form beyond the Roche limit are temporarily enveloped by vapor, as occurs to various extents in all models of the giant impact (e.g. Nakajima and Stevenson 2014).

**4.5 Accretion of the Moon**

Initial *N*-body simulations of the Moon's accumulation after a giant impact assumed a disk comprised of condensed particles (Ida et al. 1997; Kokubo et al. 2000; Sasaki and Hosono 2018). However, the disk would be at least partially vaporized when it formed and its subsequent collisional and viscous evolution may further vaporize at least its inner regions (e.g., Ward 2012, 2014). A hybrid model by Salmon and Canup (2012) described the disk interior to the Roche limit analytically as a uniform melt-vapor disk, while the growth of moonlets beyond the Roche limit and their dynamical interactions were modeled by an *N*-body simulation. The latter presumes an outer disk of condensates, appropriate for melt-dominated canonical-type disks, and possibly to the later evolution of initially more highly vaporized disks once their outer regions have cooled. The inner disk is assumed to spread via an instability-induced viscosity (Ward and Cameron 1978) regulated by its radiative cooling rate (Thompson and Stevenson 1988; see §4.3.2). The model includes resonant interactions between outer moonlets and the inner disk, which lead to a transfer of angular momentum from the latter to the former (e.g. Goldreich and Tremaine 1980), and for the formation of new moonlets when inner disk material expands beyond the Roche limit.

[Figure 6]

Figure 6a shows an example lunar growth history. In a few months, outer disk material accumulates into a proto-Moon containing about 40% of the Moon's mass. During this phase, resonant interactions with the proto-Moon cause the Roche-interior disk to be confined within the Roche limit. Subsequently, the Roche-interior disk spreads outward over tens of years, resulting in both an outward migration of the proto-Moon when its strong resonances are in the disk, and in the production of new moonlets as the inner disk's edge expands beyond the Roche limit. Moonlets spawned at the Roche limit then deliver the final ∼ 60% of the Moon's mass over ∼ $10^2$ yr.

In the Salmon and Canup (2012) model, the Moon's semi-major axis at the end of its accretion (∼ $6R_\oplus$) is substantially larger than in prior works (∼ $4R_\oplus$; e.g., Ida et al. 1997, Kokubo et al. 2000), resulting in decreased accretion efficiency per eqn. 3.2. At face value, this implies that disks produced in many impact scenarios may not be massive enough to yield the Moon and that more massive disks produced by large impactors (e.g., half-Earth or other synestia-forming impacts) may be needed. However, approximations in the model may have affected the calculated efficiency, and more accurate descriptions are needed that, e.g., resolve the inner disk's radial structure (e.g., Salmon and Canup 2019). The model also assumed that the inner disk vapor and melt components co-evolve, when instead the disk could adopt a layered structure (e.g., Ward 2012, 2017).

A different approach tailored to describing the Moon's formation from a highly vaporized synestia was developed by Lock et al. (2018). Initial condensates whose specific angular momenta exceed that of a circular orbit at the Roche limit are assumed to rapidly accumulate into a single proto-moon on a circular orbit. A radiative cooling model is then used to estimate the moon's subsequent growth from additional condensates that are produced at the low pressures of the photosphere and fall into the hotter, higher-pressure vapor in the interior of the synestia. Small condensates are rapidly re-vaporized as virga, but larger moonlets can survive for tens of years against vaporization without substantial mass loss and surrounded by tens of bars of vapor. Moonlet surface temperatures are buffered by the onset of silicate vaporization at 3000 to 4000 K (see §4.6). As the synestia cools its rotational profile adjusts, and the boundary between the corotating and disk-like region evolves. The vaporized structure retreats to within the lunar orbit, leaving a Moon near the Roche limit in less than one year. The Lock et al. (2018) model includes a variety of physical processes not considered in prior accretion models, e.g., radial pressure support, condensation, and droplet vertical and radial transport. However, the model does not directly calculate lunar accretion and neglects several physical processes, including the inner disk-moonlet interaction as well as moon(let)-gas interactions, in particular gas drag and resonant torques that could lead to radial evolution.

An impact-generated disk with sufficient mass to yield the Moon would eventually yield a single Moon (Canup et al. 1999; Crida and Charnoz 2012), although two-moon outcomes that later destabilize through, e.g., a mutual collision are possible (e.g., Ida et al. 1997; Salmon and Canup 2012).

**4.6 Volatile loss and retention after the impact**

The Moon's depletion in volatiles has often been associated with the Moon-forming impact and its formation from a partially to nearly fully vaporized disk (e.g., Nakajima and Stevenson 2014). A key volatile of interest is water. Desch and Taylor (2013) suggest that the protolunar disk's vapor phase would have been dominated by H, dissociated from $H_2O$ ($H_2O = 2H + O$) at high temperatures. Under this condition, both H and O could have hydrodynamically escaped to space, potentially entraining other heavier volatiles as well. If efficient, this process would have produced a water-poor Moon. However, bulk lunar water abundances as large as a few hundred ppm have been estimated (see §2.1 and also McCubbin et al., 202X, this volume), and indigenous lunar D/H ratios appear similar to terrestrial values (Saal et al. 2013), which seems inconsistent with significant escape of H that would have elevated lunar D/H.

Nakajima and Stevenson (2018) considered that the upper disk regions (~3-5$R_\oplus$ above the mid-plane), where escape would have occurred, would have been dominated by heavy elements/species, such as SiO and O, shortly after the impact when disk temperatures were still high (i.e., > 2000K). As the disk cooled to temperatures ≤ 1600 K, the upper regions become dominated by molecular $H_2O$ (e.g., Figs. 2-3 in Nakajima and Stevenson 2018; see also Visscher and Fegley 2013). Thus, H is predicted to be a minor species throughout the disk evolution. In this case, escape of H would require it to diffuse out from a vapor with a much higher mean molecular weight. Because diffusion-limited escape is inefficient, escape of hydrogen and other heavier volatiles, such as K, Na and Zn, would then have been minor.

How can we explain the lunar volatile depletion if escape to space was limited? A potential solution considered by several works is that disk volatiles may have been preferentially accreted by the Earth rather than by the Moon. Canup et al. (2015) combined output of lunar accretion models by Salmon and Canup (2012) with estimates for the disk temperature at the Roche limit vs. the condensation temperatures for K, Na, and Zn at this distance. Fig. 6b-c shows results for the accretion model shown in Fig. 6a. Initially, Roche limit clumps are hot and volatile-poor, and they are efficiently accreted by the Moon to form a volatile-depleted outer layer up to 500-km in depth. With time, two changes occur: (1) inner disk temperatures decrease, allowing moderately volatile elements to condense near the Roche limit, and (2) there is a dynamical transition from a regime in which clumps produced near the Roche limit are efficiently accreted by the Moon, to one in which clumps are instead scattered by the Moon back into the disk and ultimately onto the Earth. Because both processes are regulated by the evolution of the inner disk's surface density, their relative timing is coupled, and (2) occurs prior to (1) (Fig. 6b-c). This model implies that while the first portion of the Moon that accretes from outer disk material may have been volatile-rich, it would have been overlain by volatile-poor material delivered later from the inner disk. The Moon's observed volatile depletions would then reflect the relative proportions of the Moon delivered by each reservoir, as well as the extent of internal mixing in the Moon after its assembly. Greater observed depletions would result for incomplete mixing, perhaps consistent with an initial Moon that was sub-solidus beneath its magma ocean (e.g., Andrews-Hanna et al. 2013). There is also some evidence for distinct volatile and isotopic reservoirs within the Moon that vary with depth (e.g., Robinson and Taylor 2014; Cano et al. 2020). During this process, some water could have been dissolved in the silicate liquid due to enhanced solubility at the higher pressures in the mid-plane (~1-10 MPa) (Pahlevan et al. 2016; Hauri et al. 2017), which could lead to a "wet" but volatile-depleted Moon.

A different view considers that the disk vapor was much more viscous than the melt, due to, e.g., magnetic coupling. In this case the disk's volatile-rich vapor would have rapidly spread onto the Earth, potentially yielding a volatile-poor Moon (Charnoz and Michaut 2015; Nie and Dauphas 2019). Further, the rate of vapor removal could be sufficient to sustain evaporation into a vapor medium that was ~ 99% saturated, providing conditions consistent with the observed heavy isotopic compositions for moderately volatile elements in lunar rocks (Nie and Dauphas 2019; see §2.4). However, an issue with this general type of scenario is that highly viscous vapor would also have spread radially outward (e.g., Gammie et al. 2016), and it is not clear how the Moon would avoid accreting this volatile-rich (and isotopically light) material when the vapor later cooled and condensed.

Lock et al. (2018) proposed that the Moon acquired its depletion in volatiles by condensing in melt-vapor equilibrium in a local atmospheric pressure of a few tens of bar. For these conditions, there is good agreement between the Moon's bulk composition and that predicted for many elements (Lock et al. 2018, their Fig. 14). As the synestia cools, more of the vapor condenses and the pressure gradient term in equation 4.1 is reduced, causing the synestia to contract. Calculations of synestia cooling find that residual vapor in the disk-like region has limited angular momentum at the time the Moon separates from the synestia (Lock et al. 2018). The volatile elements will condense at lower temperatures (i.e., late in the sequence) at which point the proto-Moon and vapor disk may be decoupled. The amount of mass accreted to the Moon after separation from the synestia could then be minimal, with volatile-rich vapor eventually accreted by the Earth (Lock et al. 2018).

**4.7 Angular momentum transfer mechanisms**

The principal driver of the Earth-Moon system's dynamical evolution has been tides raised by the Moon on Earth. As Earth rotates faster than the Moon moves in its orbit, tidal bulges raised by the Moon in our oceans and solid Earth form ahead of the nearside sub-lunar point, causing torques between the Moon and the tidal bulges it produced to remove angular momentum (AM) from Earth's rotation and pass it to the Moon's orbital motion. This process conserves the combined AM of Earth's rotation and the Moon's orbit (e.g., Murray and Dermott 1999).

Other processes could make small changes to the Earth-Moon AM over the system's history. If a planet has a liquid core and a mantle that behaves rigidly on short-timescales, core-mantle friction arising from differing response to torques from the Sun and a satellite can reduce the planet's obliquity and spin rate (Rochester 1976), but this is most important for slowly rotating planets like Venus (Goldreich and Peale 1970). Climate friction relies on the planet's oblateness being a time-delayed function of obliquity, which may be applicable during glaciations (Williams et al. 1998), but it is very unlikely that it led to large-scale changes in Earth's obliquity or AM (Levrard and Laskar 2003). Late accretion onto Earth after the Moon formed may have been dominated in mass by large projectiles, and if so, they are estimated to have altered (increased or decreased) the Earth-Moon AM by a few percent via direct impacts (Bottke et al. 2010) and a few tens of percent via collisionless gravitational encounters (Pahlevan and Morbidelli 2015). Because Earth rotates in the same sense as its orbit with a day that is shorter than the year, the tide raised by the Sun on the Earth acts to slow Earth's spin, reducing the Earth-Moon system's AM by about 1% over geologic time (Canup 2004a,b; Peale and Canup 2015). Lunisolar cross-tides refer to the lunar interactions with the tidal bulge raised by the Sun and vice versa. The effect of the solar tidal bulge is akin to a ring around Earth that is rotated out of the ecliptic due to tidal lag (Neron de Surgy and Laskar 1997). The torque between this asymmetric ring and the Moon (which is, on average, in the ecliptic plane) acts to lower Earth's obliquity, while preserving the ecliptic component of the AM (e.g., Touma and Wisdom 1994). However, cross-tides also do not significantly affect the Earth-Moon AM.

Because such modifications are expected to be small, it had been assumed that the total AM of Earth-Moon system just after the Moon-forming impact should be comparable (i.e., to within approximately 10 to 20%) of that in the current system. However, in the past decade three effects have been proposed that could reduce the Earth-Moon system AM by much larger proportions. Two have been quantitatively modeled and involve perturbations by the Sun; a third is the suggestion that there could have been a magnetically driven outflow akin to that which can occur for young stars. We describe the basic mechanisms here; details are further discussed in §5.3.

*4.7.1 Solar Evection Resonance* As the Moon tidally evolved outward, it would have encountered the evection resonance when the lunar perigee precession period was equal to one year, occurring when the Moon was within ten Earth radii (e.g., Kaula and Yoder 1976). If the Moon is captured into evection, its outward orbital evolution is initially accompanied by a growth in orbital eccentricity, as required to maintain the precession period at one year. As large eccentricities are reached, tidal dissipation within the Moon becomes important, and this can arrest further eccentricity growth and outward tidal migration, locking the Moon at a nearly constant distance from Earth. However, such an equilibrium is not static, as Earth's rotation keeps being decelerated by tides raised on the Earth.

The location of the evection resonance is dictated by Earth's oblateness (which forces the lunar orbit to precess), in turn a function of the planet's spin rate. Thus as Earth spins down the evection resonance slowly moves inward. During this process, AM is transferred from Earth's spin to the Moon's orbit by tides on Earth, while solar resonant perturbations on the Moon remove AM from the lunar orbit (by increasing its eccentricity at an approximately constant semimajor axis) and transfer it to Earth's heliocentric orbit.

Touma and Wisdom (1998) considered an initial Earth-Moon system having the present AM, and found that while the Moon can reach high eccentricities in the evection resonance, it quickly exits the resonance with little AM loss ($< 0.1\ L_{EM}$). Ćuk and Stewart (2012) assumed a faster spinning primordial Earth, with a spin AM of 2 to $3L_{EM}$. They found that this enables a much longer capture in the resonance, during which the Earth-Moon system can lose half or more of its AM. For a range of relative tidal parameters in the Earth and Moon as expressed by the tidal "*A*" parameter (see §5.3), the AM loss ended when approximately the present Earth-Moon system AM was reached, independent of the initial AM when the Moon formed.

Ćuk and Stewart (2012) considered a relatively non-dissipative Moon, leading to high lunar eccentricities during the evection encounter. Wisdom and Tian (2015) and Tian et al. (2017) instead found that large-scale AM removal required a low-eccentricity evolution with a more dissipative Moon. In their successful histories, the Moon is not captured into evection proper, but strong satellite tides keep it just inside the evection resonance, transferring AM to Earth's heliocentric orbit through a limit cycle. Rufu and Canup (2020) found that a wide range of final system AM are possible, depending on when escape from the evection limit cycle/quasi-resonance occurs.

*4.7.2 Laplace Plane Instability* In classical tidal evolution models (e.g., Goldreich 1966; Touma and Wisdom 1994) and the evection resonance scenario of Ćuk and Stewart (2012), the lunar inclination and Earth's obliquity are assumed to have been relatively low. The post-impact obliquity of Earth is about 10° in this standard picture, gradually increasing to today's 23° during lunar recession, while the lunar inclination decreased from values as large as 12° (if the Moon were inclined when it was close to Earth) to the present value of 5° relative to the ecliptic.

However, this standard picture ignores lunar obliquity tides, which may have been strong when the lunar obliquity was high. High forced (i.e., induced) obliquities are expected at lunar separations of 30 to 35$R_\oplus$ during the "Cassini State Transition" (CST; Ward 1975) that occurs as the Moon transitions between Cassini states, which are equilibria spin axis configurations in which the timescales for lunar orbital and spin precession are equal, and the spin axis of the Moon, the normal to the lunar orbit, and the normal to the Laplacian plane are co-planar. Chen and Nimmo (2013, 2016) found that obliquity tides greatly reduce any lunar inclination present if the Moon had a magma ocean when its obliquity was significant. Ćuk et al. (2016) assumed a solid Moon at the time of the CST, but also found significant damping of lunar inclination by obliquity tides.

Explaining the Moon's current orbital tilt then requires either a much larger pre-CST lunar inclination of about 30°, or that the lunar inclination was generated after the CST. Ćuk et al. (2016) argued that the only plausible mechanism for creating a large pre-CST inclination is the tidal evolution of the Moon away from an initially high-obliquity (> 65°) Earth. For this case, when the Moon's orbit expands to a distance of about 15 to 20$R_\oplus$ it encounters an instability at the so-called Laplace plane transition, where perturbations from Earth's oblateness and the Sun compete for dominance. Chaotic lunar orbits at this distance induce significant lunar orbital eccentricity, strong satellite tides, AM transfer to Earth's heliocentric orbit, and excitation of the lunar inclination. Ćuk et al. proposed such a history as a way to both excite a large initial lunar inclination and to remove excess angular momentum after a high-AM giant impact. However, Tian and Wisdom (2020) argued that the needed initial high-obliquity state is inconsistent with the current Earth-Moon AM and Earth's present obliquity (see §5.3). Alternatively, Pahlevan and Morbidelli (2015) proposed that the lunar inclination is a result of multiple scattering encounters between the Moon and large lunar-sized bodies left over after the Moon-forming impact. Depending on the mass in these bodies and the rate of the Moon's orbital expansion, such scattering may allow the lunar inclination to be generated after the CST.

***4.7.3 Magnetic Coupling.*** Gammie et al. (2016) proposed that the Earth-Moon system could lose angular momentum through a magnetically driven outflow originating within the boundary layer between the outer post-impact Earth and the proto-lunar disk. This process would require a rapidly rotating Earth, a strong magnetic field in the boundary layer and the disk, and an ionized wind that could couple to the field (Gammie et al. 2016).

## 5. FACTORS THAT INFLUENCE MOON ORIGIN SCENARIO LIKELIHOODS

The famous principle of Occam asserts that the simplest explanation for an observation is preferred, or alternatively, that "more things should not be used than are necessary". Giant impact models initially sought to explain the Earth-Moon system through a single event, assuming that (*i*) one collision produced the Moon, (*ii*) the angular momentum of the current Earth-Moon system was due to the Moon-forming impact, (*iii*) the Moon's composition reflected the composition of the disk produced by the impact, and (*iv*) the impactor's composition was likely as different from Earth's as Mars, the only other planet whose isotopic composition is currently known. Work to date implies that, taken as a group, assumptions (*i*) through (*iv*) appear inconsistent with the Earth-Moon system we see.

One possible explanation is that the Moon did not form by impact. However, it seems very difficult to explain basic characteristics, such as the Moon's lack of iron, by intact capture or co-accretion[1]. Alternatively, one or more of assumptions (*i*) through (*iv*) may be incorrect. A common early approach relaxed assumption (*iv*), arguing that the impactor would have been isotopically Earth-like if it formed near 1 AU (e.g., Wiechert et al. 2001; Belbruno and Gott 2005). This seemed plausible because canonical impacts required low relative velocities at infinity, consistent with an Earth-like impactor orbit. However, soon after *New Views of the Moon* (2006), the "canonical impact + Earth-like impactor" concept was challenged by Pahlevan and Stevenson (2007), who used impact statistics from an *N*-body planet accretion simulation by Chambers (2001) to estimate that the probability of an isotopically Earth-like giant impactor was of order a percent or less. This was viewed as disturbingly low, which provided a primary motivation for developing alternative concepts that might more naturally account for both the dynamical and compositional properties of the Earth-Moon. The subsequent decade witnessed tremendous new model development, starting with the equilibration model proposed in Pahlevan and Stevenson (2007) and leading to a variety of scenarios highlighted throughout this chapter.

In this section, we reconsider the issue of probability in light of such recent works. We begin by scrutinizing several overarching issues of key importance to current origin models: the likelihood of an impactor that was Earth-like in isotopic composition (§5.1), the potential for Earth-protolunar disk compositional equilibration to remove compositional differences and yield an isotopically similar Earth and Moon (§5.2), and constraints on large-scale angular momentum modification after the Moon-forming impact (§5.3). In §5.4, we review primary conditions needed for success in each of current impact scenarios.

---

[1] Fission is now plausible, given the identification of mechanisms capable of large-scale modification of the Earth-Moon angular momentum after the Moon formed (indeed, the Ćuk & Stewart (2012) model is akin to an impact-induced fission). Fission models could explain the Earth-Moon isotopic similarities, but would not explain the lunar volatile depletion without an additional process(es).

**5.1 Likelihood of an Earth-like composition impactor**

Most impacts produce disks derived substantially from the impactor. How likely is it that the impactor would have had an Earth-like composition? We first consider oxygen as a proxy for elements whose isotopic compositions in the solar nebula may have varied with radial distance from the Sun. Assuming an initial radial gradient in oxygen isotopic composition consistent with the Earth-Mars difference ($\Delta^{17}O \approx 0.32‰$), one can use $N$-body planet accretion simulations to estimate the likelihood, $P_{match,O}$, of an impactor having an Earth-like O isotopic composition by virtue of having formed near 1 AU (e.g., Pahlevan and Stevenson 2007). Recent works predict that the probability of a final impactor with $|\Delta^{17}O| \leq 0.015‰$ is $P_{match,O} \approx 5$ to 10% (Kaib & Cowan 2015a,b; Mastrobuono-Battisti et al. 2015, Nakajima and Stevenson 2015; Mastrobuono-Battisti and Perets 2017), depending, e.g., on how one defines the initial gradient and what constitutes an Earth-analog and its final giant impactor.[2]

Alternatively, the difference in oxygen composition between Earth and Mars may *not* be representative of differences among Earth's final large impactors. By virtue of its much larger mass, Venus would be a better indicator of overall inner solar system composition (Canup 2013), however its isotopic composition remains unknown. Perhaps relatively tiny Mars is a compositional outlier and the rest of the inner Solar System accreted from a more isotopically uniform reservoir. Several arguments support this view. First, if the isotopic compositions of many elements had varied across the inner nebula, each with its own radial distribution due to mass independent fractionation processes specific to the particular element, it would seem extremely improbable to find among our meteoritic collection some samples like the enstatite meteorites that are isotopically similar to the Earth and Moon across many elements (e.g., O, Ca, Ti, Cr, Mo and Ru; Dauphas et al. 2014). Second, lithophile vs. moderately or highly siderophile elements in Earth's mantle reflect different stages of Earth's accretion, and can thus be used to probe how the composition of accreted material changed with time (Dauphas 2017). Such analysis implies that the final 40% of Earth's mass was accreted from material isotopically indistinguishable from enstatite meteorites (Dauphas 2017; see also Bermingham et al. 2018). In the limit that Earth's late impactors had an enstatite-like compositional dispersion with $\Delta^{17}O \approx -0.01 \pm 0.06$ (e.g., Dauphas 2017), the probability of an impactor with $|\Delta^{17}O| \leq 0.015‰$ would be $P_{match,O} \approx 0.2$ based on the $N$-body simulations referenced in the prior paragraph. Together these estimates and those in the prior paragraph suggest $P_{match,O} \approx 0.05$ to 0.2, a more optimistic appraisal than in Pahlevan and Stevenson (2007).

However, explaining the Earth-Moon similarity in tungsten isotopes imposes an additional and separate constraint. The W isotopic composition of a differentiated body's mantle is set by the conditions and timing of its core formation. These would likely differ for the protoearth and impactor, even if their other isotopic compositions (e.g., O) were similar. To explain the matching $\varepsilon^{182}$W values inferred for the initial lunar and terrestrial mantles (e.g., Touboul et al. 2015) the impactor and protoearth $\varepsilon^{182}$W values and Hf/W ratios must assume specific values. Reasonable histories yield these for a Mars-sized impactor (Dauphas et al. 2014a), but such a match is improbable. Kruijer and Kleine (2017) considered plausible fractions of impactor mantle and core, and protoearth mantle in the post-impact Earth and Moon, together with varied values for, e.g., partition coefficients, core formation timescale, and degree of impactor core/protoearth mantle equilibration during the Moon-forming event. They estimate the probability of producing a planet and moon as similar in W as the inferred initial Earth and Moon is $P_{match,W} \leq 5\%$, even in the limit that the disk's silicate was derived entirely from the protoearth's mantle. Fischer et al. (2018, 2020) combined $N$-body planet accretion simulations with a core differentiation model, and estimated a < 2 to 5% probability that Earth and the Moon-forming impactor had the needed W compositions; a similar approach using Grand Tack simulations yielded a $\leq 8\%$ probability (Zube et al. 2017). Overall these works estimate $P_{match,W} \approx 0.01$ to $0.1$. This implies an impactor "match" in both O and W would have a likelihood of $\approx$ a percent or less, a comparable or even more restrictive assessment than found in Pahlevan and Stevenson (2007) who considered only oxygen.

---

[2] We know of no N-body accretion simulations in which a Moon-forming impactor accretes from debris at Earth's L4 or L5 Lagrange points as proposed by Belbruno and Gott (2005; see also Kortenkamp & Hartmann 2016).

## 5.2 Potential for compositional equilibration between the protoearth and protolunar material

Consider an impact that produces a protolunar disk with an isotopic composition distinct from that of the protoearth's mantle. Portions of the post-impact system would seem likely to mix due to convection and/or other processes (§4.3; Pahlevan and Stevenson 2007; Lock et al. 2018). What is the potential for such mixing to be sufficient to yield compositional equilibration and an isotopically similar Earth and Moon?

A first requirement is that mixing occurs before the Moon forms. If the rate limiting step is radial mixing in the disk, a diffusivity $D = \alpha cH$ (where $\alpha$ is a constant, $c$ is the gas sound speed, and $H$ is the pressure scale height of the disk atmosphere) could mix the disk in a few hundred years or less if $\alpha \sim$ few $\times 10^{-4}$ to $10^{-3}$, consistent with values estimated for convective turbulent motions (Pahlevan and Stevenson 2007). The inner (Roche-interior) disk would take decades or longer to cool and condense, allowing time for it to potentially mix with the Earth. However, outer (Roche-exterior) disk material may accumulate into a large moonlet in less than a year (Salmon and Canup 2012; Lock et al. 2018). If outer disk material did not equilibrate, a successful outcome would be possible if inner disk material with an equilibrated composition was the last material accreted by the Moon, producing an equilibrated outer layer reflected by our lunar samples (Salmon and Canup 2012). This would require that the Moon derived a substantial fraction of its mass from the inner disk, and that the Moon's interior did not fully mix during or after its accretion, which are plausible but restrictive conditions.

For thorough equilibration, the process that drives mixing must not also produce outward angular momentum transport with comparable efficiency, because if it did, much of the disk mass would be driven inward on the same timescale (Stevenson 1990; Melosh 2014), limiting the magnitude of the mixing. Thus, disk turbulence produced by the magnetorotational instability may be a poor candidate to drive mixing, because it would cause substantial mass accretion onto Earth before significant mixing took place (e.g. Carballido et al. 2005, Gammie et al. 2016). Thermal convection in the disk is more promising, because it may be inefficient in transporting angular momentum in differentially rotating flows (Held and Latter 2018). However, the total mass exchange needed to remove a large Earth-disk compositional difference across the radial extent of the inner disk is greater than the inner disk's total mass, and whether this degree of exchange could occur without substantial angular momentum transfer is unclear (Melosh 2014).

Immediately after a Moon-forming impact, the Earth's mantle has an inverted thermal structure, with its outer, more impactor-rich layers having the highest temperatures and entropies (e.g., Canup 2004a, her Fig. 4b; Nakajima and Stevenson 2015, their Figs. 2 and 3; Lock et al. 2018, their Fig. 15; Deng et al. 2019). These outer layers would need to cool before the bulk mantle could undergo thermally-driven convection (Pahlevan et al. 2011), which may take longer than the time for forming the Moon (Nakajima and Stevenson 2015). Perhaps if the chemical composition of the impactor mantle was distinct from that of the BSE (e.g., was more FeO-rich, similar to Mars), the negative chemical buoyancy of the hot, impactor-rich shell could have driven magma ocean overturn and mixing. Alternatively, if the Earth's mantle did not fully mix prior to the Moon's assembly, protolunar material might equilibrate only with the planet's outer high-entropy, impactor-rich regions (Nakajima and Stevenson 2015; Lock et al. 2018). Such limited mixing would be less effective in equalizing the disk composition with that of the full BSE, but it might still be sufficient to dilute modest disk-planet compositional differences while also allowing for the preservation of isotopically distinct regions in the Earth's mantle (see also §4.1). Alternatively, it has been suggested that perhaps the Earth's mantle never fully mixed over its entire history (see §4.1), so that knowledge of the BSE composition is biased by the upper, impactor-dominated mantle layers preserved since the giant impact (whose composition is similar to that of the Moon), with an isotopically distinct lower mantle beneath (Deng et al. 2019).

High-AM/high-energy impacts produce highly-vaporized synestias that may be favorable for at least partial equilibration, particularly within the co-rotating region that lacks a radial gradient in angular velocity that could impede radial mass transport/mixing (e.g., Lock et al. 2018, their fig. 3D). However, the outer edge of the co-rotating region (e.g., ~ $1.6R_\oplus$ for an Earth-like body with an angular momentum of 2 to $2.5L_{EM}$; see Fig. 4.2) is well inside the Roche limit ($2.9R_\oplus$) and the distance at which the Moon forms. Thus, even if the co-rotating region of a synestia were well-mixed, equilibration of Earth-Moon compositions would additionally require mixing across large radial distances in the outer, differentially-rotating disk-like region, for which the constraint discussed above that mixing must occur on a timescale shorter than that associated with angular momentum transport (Stevenson 1990; Melosh 2014) would still apply. Whether thorough equilibration would be expected throughout the co-rotating region is unclear, as the lower entropy, inner portions of the post-impact Earth may be unlikely to mix with the higher entropy outer regions, as discussed in the prior paragraph.

Given the complexity of the physics of equilibration, independent markers of its importance in establishing the Moon's bulk composition have been sought. Differences between the Earth and Moon isotopic compositions for highly refractory elements would seem to provide evidence for equilibration, because at least in vertically stratified disks (Ward 2012, 2014, 2017), mixing would be restricted to elements in the vapor phase. Instead, the Earth and Moon have indistinguishable Ti compositions (Zhang et al. 2012). This does not, however, rule out equilibration, because Ti may have been partially vaporized (Zhang et al. 2012), or refractory-rich droplets could have equilibrated with surrounding vapor if the disk was vertically unstratified (Pahlevan et al. 2011). Recent high precision Ca data nominally imply a difference between the Earth and Moon Ca isotopic compositions, but it is barely resolvable (< 10 ppm; Schiller et al. 2018).

Pahlevan et al. (2011, their Fig. 6) predicted that equilibration between the disk and the full terrestrial mantle would lead to a relationship between resulting Earth-Moon differences in FeO/MgO and in silicon isotopic composition, with larger Earth-Moon differences in FeO/MgO implying larger differences in $\delta^{30}$Si. In the framework of this model, the fact that the Earth and Moon are now known to have essentially identical Si compositions ($\Delta^{30}Si_{BSE-lunar} \equiv \delta^{30}Si_{BSE} - \delta^{30}Si_{lunar} = 0.00 \pm 0.022$‰; Armytage et al. 2012) would imply that they should have similar FeO/MgO as well, i.e., that the lunar Fe/(Fe + Mg) ratio must be between 1 to 1.3 times that of the BSE (Armytage et al. 2012). Some prior estimates fell in this range (~ 1.2, Warren 2005; see also Table 1 of Taylor et al. 2006b), but more recent work including GRAIL results estimate this ratio is 1.4 (Warren and Dauphas 2014; Dauphas et al. 2014a; see also Sakai et al. 2014), nominally inconsistent with equilibration. However, there are still substantial uncertainties that could permit a more BSE-like value too.

Thus, equilibration is an appealing process that could account for a wide range of Earth-Moon isotopic similarities, including both O and W. It is expected that at least portions of the protolunar material would have diffusively mixed and equilibrated with at least portions of the Earth's mantle. However, to date we lack rigorous mixing models that include disk/planet cooling and Moon accumulation. Thorough equilibration could lead to predictable chemical signatures, but we lack clear evidence of its occurrence, and there is some ambiguous contrary evidence. Thus, the likelihood of equilibration sufficient to equalize Earth-Moon isotopic compositions remains highly uncertain.

### 5.3 Constraints on large-scale modification of the Earth-Moon angular momentum

Imagine the initial Earth-Moon system had a significant angular momentum excess (*i.e.*, ≥ 1.3$L_{EM}$) compared with the current Earth and Moon. Three mechanisms have been proposed that could extract substantial AM and potentially yield an appropriate final system value (§4.7).

In the models of Ćuk and Stewart (2012), the requirement that the Moon's encounter with the evection resonance with the Sun produces a final system whose angular momentum is $L_F \sim L_{EM}$ constrains the relative strength of tidal dissipation in the Moon and Earth, expressed by the parameter $A \equiv (k_{2M}/k_{2\oplus})(Q_\oplus/Q_M)(M_\oplus/M_M)^2(R_M/R_\oplus)^5 \approx 10 \ (k_{2M}/k_{2\oplus})(Q_\oplus/Q_M)$, where $k_2$ is the Love number, $Q$ is the tidal dissipation factor, and subscripts $\oplus$ and $M$ correspond to terrestrial and lunar values. Successful cases in Ćuk and Stewart (2012) had $0.8 \leq A \leq 1.7$, similar to $A$ for the current Earth-Moon ($A \approx 0.5$).

For such low-$A$ histories, evection produces high lunar eccentricities, which would lead to increased tidal dissipation in the Moon. Tian et al. (2017) included effects of eccentricity-driven tidal heating on the Moon's tidal properties, and found that the resonance is destabilized with little or no AM removal. Instead, they identified a limit cycle associated with evection (in which the Moon's orbit is regulated by evection even though the resonance is not formally occupied) that can appropriately reduce the system AM for higher $A$ values (Wisdom and Tian 2015; Tian et al. 2017). Using the Mignard tidal model (Mignard 1980), Rufu and Canup (2020; also Ward et al. 2020) predict early escape from formal resonance and also find a limit cycle/quasi-resonance that can remove substantial AM. In their models, the Earth-Moon AM after its encounter with evection (which for successful cases would need to be ~ $L_{EM}$), depends strongly on both $A$ and the Earth's tidal parameters (Rufu and Canup 2020, their figure 6).

It is probable that $A$ at the time of the Moon's initial encounter with evection would have been large, because the post-impact Earth would have been hot and fluid-like in its tidal response, implying very large $Q_\oplus$ for several million years, while the Moon would have rapidly partially cooled to a dissipative state, implying small $Q_M$ (Zahnle et al. 2015; Chen and Nimmo 2016). Simulations of the Moon's encounter with evection that adopt the Zahnle et al. (2015) model for how the Earth's tidal parameters and $A$ evolve with time yield final systems whose AM is too high for consistency with the current Earth-Moon (Rufu and Canup 2020). However, the Zahnle et al. model is simplified, and perhaps improvements to it would yield successful cases.

A second AM-reducing mechanism invokes an Earth whose obliquity was ~ 65 to 75° just after the Moon-forming impact (Ćuk et al. 2016). For a high-obliquity Earth, the Moon encounters a Laplace plane instability as it tidally evolves outward, which increases the Moon's orbital inclination and eccentricity, while decreasing the Earth's obliquity and the system AM. Tian and Wisdom (2020; also Atobe and Ida 2007) emphasize that such a history conserves the component of the Earth-Moon AM that is perpendicular to the ecliptic, and conclude that this mechanism cannot reproduce the current Earth-Moon because it leaves Earth with an obliquity that is too large and a system AM that is too high (by $\geq$ 30%), regardless of tidal model or parameters.

A third concept envisions a magnetized wind that removed angular momentum from the initial Earth-disk system (Gammie et al. 2016), but this has not been quantitatively modeled. It is unknown whether an appropriate final AM would be expected if conditions needed for an outflow were achieved.

## 5.4 Needed conditions for proposed Moon-forming impact scenarios

We here consider some of the main requirements of current impact models. When discussing constraints on impactor(s) mass and velocity, we rely primarily on results of $N$-body planet accretion simulations that adopt standard (Kaib and Cowan 2015a; Quintana et al. 2016), annulus (Kaib and Cowan 2015a), or Grand Tack (Jacobson and Morbidelli 2014) histories. Of the Quintana et al. (2016) simulations, we cite those that included a simple treatment of fragmentation. We do not include impact angle in our discussions here; each scenario generally requires a certain range in this parameter that would impose an additional constraint.

*Scenario 1: Canonical impact + Earth-like composition impactor.* $N$-body simulations suggest that a few to of order 10% of Earth-analogs experience a final giant impact with $v_i/v_{esc} <$ 1.2 and $0.1 < \gamma < 0.2$ (e.g., Raymond et al. 2009; Elser et al. 2011; Jacobsen and Morbidelli 2014; Quintana et al 2016). In a canonical scenario, the Moon-forming impact leaves the Earth-Moon system with approximately its current angular momentum. This can occur across a range of pre-impact target and impactor spin states, from non-spinning bodies to targets with pre-impact spin periods as short as 4 to 5 hr (e.g., Canup 2004a, 2008). However, a canonical model generally requires that the protoearth's spin angular momentum at the time of the Moon-forming impact was < 1 $L_{EM}$. N-body simulations by Kokubo and Genda (2010) imply that perhaps only 30% of Earth-analogs would satisfy this condition; this percentage would be much higher if rotation periods were typically ≥ 6 hr (Rufu et al. 2017).

By far the most constraining aspect of this scenario is the requirement of an isotopically Earth-like Theia. From §5.1, it appears highly unlikely that Theia would have had the proper composition to explain both the O and W isotopic similarities of the Earth and Moon.

*Scenario 2: Canonical impact + equilibration.* A canonical impact followed by equilibration would be more probable than Scenario 1 if the likelihood of equilibration between the Earth's mantle and the protolunar disk was greater than the likelihood of a needed impactor composition in both O and W. However, the likelihood of sufficient equilibration is unknown (§5.2).

*Scenario 3: Canonical impact + Earth magma ocean.* Hosono et al. (2019) found massive ($M_D \geq M_M$) disks with $|\delta f_T|$ as low as ~ 10 to 20% when they combined a hard-sphere EOS for the Earth's mantle (to mimic a surface magma ocean) with a new SPH code. This is still too large to account for the Earth-Moon O similarity for an isotopically Mars-like impactor, but would suffice if the impactor was isotopically like an enstatite chondrite. Accounting for the Earth-Moon W match would require a separate constraint on impactor composition.

A key uncertainty is the dependence of the Hosono et al. results on the use of a new SPH code, "DISPH" (Saitoh and Makino 2013). Inherent to the SPH method is "smoothing" across overlapping particles, so that a continuous flow can be described by discrete interpolation points. Standard SPH codes smooth over density, which does a poor job of handling density contrasts, *e.g.*, across a core-mantle boundary. DISPH was designed to instead smooth over pressure to mitigate such issues. However, simulations of collisions between even uniform composition bodies (without core-mantle boundaries) show different results with standard SPH vs. DISPH (Hosono et al. 2016). Furthermore, while Hosono et al. (2019) found favorable, target-dominated disks with the hard-sphere/magma ocean EOS and DISPH, this was not seen when the same EOS was used with standard SPH, where their massive disks instead had on average ~ 65% impactor material by mass (Hosono et al. 2019). Impactor-dominated disks were also found by Wissing and Hobbs (2020), who applied a different EOS to simulate impacts into a protoearth with a magma ocean using standard SPH. Grid codes such as CTH do not have the same smoothing issues as SPH, and canonical impacts modeled with both CTH and standard SPH have also yielded impactor-dominated disks (Canup et al. 2013). Additional scrutiny of the new DISPH method, in particular through direct comparisons with grid codes, is needed to substantiate the postulated magma ocean effect. Moreover, Hosono et al. (2019) used fixed surface magma ocean depths and did not consider impact-induced melting, even though it is expected to be significant (Nakajima & Stevenson 2015). A coherent thermodynamic model is also needed to validate this model.

*Scenario 4: Hit-and-run impact.* This model yields a more target-rich disk ($\delta f_T$ ~ − 50 to −35%) than a canonical impact, allowing for an impactor with $|\Delta^{17}O| \leq 0.03$ to still be consistent with the Earth-Moon oxygen similarity. In this case, $P_{\text{match,O}}$ is increased ~ 0.1 to 0.4, based on arguments in §5.1. An additional impactor compositional "match" would still be required for tungsten. The likelihood of an appropriate composition impactor is improved relative to Scenario 1, but not dramatically so. Alternatively, a hit-and-run impact could be followed by equilibration, the likelihood of which is uncertain.

A hit-and-run impact that produces a sufficiently massive disk leaves a system with $L \sim$ 1.3 to $1.4 L_{EM}$ (e.g., Canup 2014; Deng et al. 2019), so that AM removal is also required (§5.3). Here the needed degree of angular momentum extraction is less than in Scenarios 5 and 6 (and perhaps 7), but whether this would be more or less easily achieved is unclear. For an initially more slowly rotating Earth, the position of evection shifts inward, which may be less favorable for prolonged interaction with evection compared to initially higher-AM cases (e.g., Touma and Wisdom 1998; Ćuk and Stewart 2012).

*Scenario 5: Fast-spinning Earth impact*. In this model, protolunar material is derived overwhelmingly from the Earth's mantle, so that the Earth-Moon isotopic similarity in O is naturally explained. However, a constrained impactor W composition is still required (e.g., Kruijer and Kleine 2017), as well as angular momentum removal (§5.3).

An additional requirement is a protoearth rotating with a near fission rate – i.e., with a rotational angular momentum of 2 to $3 L_{EM}$ for an $\approx 1 M_\oplus$ body with an Earth-like moment of inertia – at the time of the Moon-forming impact. Kokubo and Genda (2010) use impact angular momenta from $N$-body simulations to estimate planet rotation rates assuming a planet density of 3 g/cm$^3$, so that the angular momentum of a critically rotating, Earth-mass body is ~ $4 L_{EM}$ in their analyses. They find final planet mean rotation rates that are ~ 70% of their instability rate (their Table 2 and eqn. 5). Their treatment of protoplanet collisions, while an improvement over prior works (e.g., Agnor et al. 1999), assumes that each collision results in either a perfect merger of mass and angular momentum into a single body (if the impact velocity is below a critical value), or inelastic rebound (if $v_i$ exceeds this value) and fragmentation is ignored. This parameterization will tend to overestimate planet spin, because even in largely accretionary collisions, angular momentum will be contained in material that escapes or goes into orbit around the planet (e.g., Fig. 7). However, if the effect of such processes was relatively modest, the needed fast spin could be quite probable. It would then be somewhat puzzling that such high spin planets do not exist in the inner solar system today, particularly for Mars whose spin rate is not thought to have slowed significantly over its history.

[Figure 7]

An alternate view is that the spin imparted by multiple impacts saturates at a level much less the instability rate, as may occur if the fraction of escaping angular momentum tends to increase with the planet's pre-impact spin rate (e.g., Dobrovolskis and Burns 1984; Rufu et al. 2017). In this case, a near-fission rotation would likely require a single large impact prior to the Moon-forming event. Achieving a protoearth rotation with $> 2 L_{EM}$ by a single impact requires an impactor with $\geq 40\%$ of Earth's mass ($\gamma \geq 0.4$; Figure 7). Planet accretion models predict that such collisions are uncommon at the end of Earth's growth. Out of 238 Earth-analogs in Jacobsen and Morbidelli (2014), 7% had a final giant impact with $\gamma \geq 0.4$. None of the 104 Earth-analogs in Kaib and Cowan (2015a) had final $\gamma \geq 0.4$ impacts, while 14% did in the Quintana et al. (2016). Impacts with $\gamma \geq 0.4$ that produce rapidly spinning targets also usually produce large moons (e.g., Scenario 6 below). As a large moon tidally evolved outward, planet-moon tides and/or interactions involving the Sun (evection or Laplace plane instability) would slow the planet's rotation (e.g., Canup 2014). In this view, a protoearth with a near-fission spin rate just prior to the Moon-forming impact would be very unlikely.

To trigger Moon formation from the Earth's mantle, the fast-spinning Earth model requires a relatively high-velocity, sub-Mars-to-Mars sized impactor with $v_i = 1.5$ to $3v_{esc}$ (Ćuk and Stewart 2012). This corresponds to a velocity at infinity of $v_\infty = 11$ to 28 km/sec for an ~ Earth-mass target, while the Kepler velocity at 1 AU is 30 km/sec. Thus, an impactor with a large orbital eccentricity ($e \sim 0.3$ to 0.9) and/or a high inclination is needed, perhaps requiring an outer Solar System origin (A. P. Jackson et al. 2018).

*Scenario 6: Half-Earth impact.* A relatively low-velocity ($v_i/v_{esc} < 1.6$) collision by an impactor with > 40% of Earth's mass can directly create a disk and planet with essentially identical O isotopic compositions, although a constrained range of impactor W composition is still required. Successful cases occur for many pre-impact rotation states, from nonrotating bodies to those with target pre-impact days as short as 3 hr (Canup 2012). Subsequent large-scale angular momentum extraction is required (§5.3). In addition, and as discussed in Scenario 5 above, the proportion of Earth analogs that experience a final impact with $\gamma \geq 0.4$ appears to be very small (Jacobsen and Morbidelli 2014; Kaib and Cowan 2015a; Quintana et al. 2016).

*Scenario 7: General synestia impact with equilibration.* A wide array of high-energy impacts can form synestias, and some portion of these would be massive and extended enough to produce a lunar-mass Moon. A fast-spinning Earth impact or a half-Earth impact are end-member cases of synestia-producing impacts that have been "tuned" to directly yield planets and disks with approximately equal silicate isotopic compositions. However, impacts that are intermediate to these (e.g., with $\gamma \sim 0.3$ as in Lock and Stewart 2017, their Fig. A4) could also be viable if equilibration removed compositional differences between the inner co-rotating region that includes the protoearth and the outer, differentially rotating disk-like region where the Moon forms (see §5.3). Angular momentum extraction is likely also required for a synestia-forming impact capable of producing a lunar-mass Moon.

*Scenario 8: Multiple impacts.* The multiple impact model envisions Moon formation through a series of impacts by sub-Mars mass bodies, rather than via a single larger impact. In Quintana et al. (2016, their Fig. 10), 4% of Earth-analogs grew to their final size without any "giant" impacts, defined as having a specific impact energy comparable to the high-angular momentum impacts in Ćuk and Stewart (2012) and Canup (2012), while another 16% experienced only a single giant impact. For a series of 20 sub-Mars impactors, Rufu et al. (2017) find that 40% of histories produce a lunar-mass Moon in the limit that all sequentially formed moonlets merge. However, Citron et al. (2018) find that a system of two prograde moonlets ends in a merger in less than half of their studied cases, although in ~ 70% of cases at least the outer moonlet survives.

Of the Rufu et al. impact histories with impact velocities between 1 and $2.5v_{esc}$, 13% achieve a predicted Earth-Moon match in oxygen isotopic composition. A higher success rate is associated with higher velocity impactors (2.5 to $4v_{esc}$), but those appear less probable as the planet's mass approaches an Earth mass (e.g., A. P. Jackson et al. 2018). The required high speed of the impactors may imply they formed far from 1AU, which could result in greater differences in oxygen isotopic composition than estimated based on Earth and Mars. The probability for an Earth-Moon W similarity after many moonlet-forming impacts is unclear, as this differs from the single impact scenarios studied previously. Unless the multiple impacts occurred after 60-70 Myr and/or the late growth remelted everything, a range of W anomalies could result. Finally, in only about 10% of the Rufu et al. histories does the final system angular momentum reach values comparable to $L_{EM}$, although about 50% of cases have a system with $\approx L_{EM}$ at some point in their history.

In the multiple impact scenario, the Moon grows alongside the Earth as the latter accretes perhaps 50% of its mass. The growing Moon would then be vulnerable to its own interactions with a massive population of background planetesimals, which, depending on their size distribution, could overly pollute the Moon with siderophile-rich material (e.g., Canup and Asphaug 2001), erode it via high-velocity impacts (Raymond et al. 2013), or dislodge it from Earth orbit by collisionless scattering (Pahlevan and Morbidelli 2015; Rufu et al. 2017).

## 5.5 Discussion

Decades of modeling have shown that large impacts are efficient producers of moons. However, issues discussed in this section suggest that the overall likelihood of explaining the particular characteristics of our Earth-Moon system may be small, even given innovative and diverse impact models. We expect each impact-generated planet-moon system to be different, and so perhaps finding that our Earth-Moon may reflect a rare outcome compared to all possible systems should not be surprising. As observed by Melosh (2014), "How special a set of initial conditions should we find acceptable? Unfortunately, we have only one Moon, and so one can always argue that we are simply very, very "lucky" to have one whose isotopic ratios are so similar to the Earth." Future work could rule out most of the models considered here, and then (in the spirit of Sherlock Holmes) whatever left, however improbable, will be the solution. "Nevertheless," cautioned Melosh (2014), "good science suggests that we should not rest on one (or two!) possible but unlikely solution(s), but continue to seek for one whose probability of occurrence is much higher." Perhaps a process that at this time appears constraining may later be understood to be probable. Or perhaps there is a more probable solution that eludes us still.

## 6. CONCLUSIONS AND OPEN ISSUES

The Moon's origin is coupled to that of the Earth. Because of the Moon's accessibility, our knowledge of its detailed composition and physical properties will likely always exceed that of the other planets in the inner Solar System. Thus, unraveling how the Moon formed holds the promise of providing a cornerstone for understanding how our terrestrial planets originated. Realizing that promise has proved challenging. Current models provide an array of possible solutions, but all involve conditions whose likelihood appears small or remains highly uncertain. Particular topics and/or datasets that could help reduce current uncertainties include:

- *Isotopic composition of Venus*. Current knowledge of the isotopic composition of Mars heavily influences our thinking about the compositional diversity of the inner solar nebula.

Knowledge of the isotopic composition of key elements in Venus, a much more massive planet, would provide a stronger constraint. If Venus proved Earth-like, the likelihood of an Earth-like impactor would be increased (Canup 2013; Righter 2007; Greenwood and Anand 2020).

- *The potential for planet-disk equilibration and theoretical, numerical and/or observational tests for or against equilibration.* Quantitative models of the physics of planet-disk mixing – especially to characterize the degree of disk radial mixing — are needed; additional high-precision measurements of the lunar and terrestrial Ca isotopic compositions would also be valuable.

- *The feasibility of substantial modification of the Earth-Moon angular momentum.* Many current impact models require large angular momentum (AM) extraction from the initial Earth-Moon system. Interaction of the Moon with the evection resonance can enable the needed angular momentum extraction for limited parameter regimes (Wisdom and Tian 2015; Tian et al. 2017; Rufu and Canup 2020), but whether this is likely for realistic models of the post-impact Earth and Moon and/or for various initial system angular momenta remains unclear.

- *Origin of the Moon's orbital inclination.* Explaining the Moon's current 5° tilt relative to the ecliptic has been a longstanding issue for a giant impact origin, which would nominally produce a low-inclination Moon (< 1°). Proposed solutions to the "lunar inclination problem" invoke one or more subsequent excitation processes, including the evection and eviction resonances (Touma and Wisdom 1998), Moon-protolunar disk resonant interactions (Ward and Canup 2000), planetesimal scattering (Pahlevan and Morbidelli 2015), or a Laplace plane instability (Ćuk et al. 2016). Recently it has been recognized that inclination damping, as well as excitation, may have been important during the Moon's tidal evolution. Large-scale inclination damping may have occurred at an Earth-Moon separation of about 30 Earth radii when the lunar obliquity and associated dissipation in the Moon increased (Chen and Nimmo 2016; Ćuk et al. 2016). That the Moon's orbit is inclined today suggests that either the Moon acquired its current inclination later in its evolution after such damping, perhaps via late scatterings (Pahlevan and Morbidelli 2015; Chen and Nimmo 2016; although see also Ćuk et al. 2016), or that the Moon's initial inclination was very large so that even after strong damping the modern 5° tilt might remain (Ćuk et al. 2016; although see also Tian and Wisdom 2020). Further work addressing this issue could provide new constraints on origin models.

- *Additional constraints on the Moon's initial thermal state*. A lunar seismic network, as well as additional thermal models incorporating GRAIL data, could reveal new constraints on the depth of the initial lunar magma ocean. The abundance of light elements in the core would provide key information of the lunar accretion process. If, as argued by Charlier et al. (2018), the Moon cannot have been initially fully molten, this would provide a stringent constraint on lunar origin models (e.g., Pritchard and Stevenson 2000), potentially ruling out those in which the Moon forms in << 100 yr (Barr 2016).

- *Evolution of planet rotation and moon formation through multiple impacts.* Many accretionary impacts will produce moons, whose subsequent evolution will affect planet rotation rates and the potential for moon-moon mergers. Exploration of how multiple impacts contribute to planet rotational angular momentum evolution and consecutive moon formation and loss is warranted to better understand how multiple impacts may have contributed to the final Earth-Moon.

- *The Moon's volatile content and its origin.* Additional analyses of lunar samples and acquisition of future samples that may further reflect the Moon's composition at depth (e.g., glasses or samples of the Moon's mantle potentially exposed at large impact basins such as South Pole Aitken) could place improved constraints on the thermal evolution of the protolunar material and whether the Moon's volatile content increases with depth (e.g., Robinson and Taylor 2014). The Moon's volatile content relative to the Earth is a fundamental constraint, and further investigation beyond the few, rather schematic models developed to date is needed to account for both the Moon's relative depletion in volatiles and its isotopically heavy composition in at least some volatile elements.

- *Improved models of the disk's evolution and Moon assembly.* There remains substantial uncertainty as to whether disks produced in most giant impact models are actually massive enough to produce the Moon, given recent accretion efficiency estimates (Salmon and Canup 2012). How a part melt, part vapor disk evolves into the Moon is a challenging problem — in which disk dynamics and thermodynamics are fundamentally linked – that merits further investigation.

- *Further assessments of the potential role of magnetic fields in disk evolution.* If the protolunar disk was subject to magnetic-induced turbulence, rapid radial spreading could result in a different evolution than considered in nearly all prior works (Gammie et al. 2016). Key open issues include the expected ionization state in a two-phase disk undergoing condensation rain-out and the presence and expected strength of an initial "seed" field in the post-impact environment.

- *Coupled dynamical-compositional models to better address the origin of enstatite chondrites.* New evidence suggests that the Earth's building-blocks were enstatite-like in isotopic composition (Dauphas 2017). How this arose in the same general environment that produced substantial Earth-Mars isotopic differences and a broad range in isotopic compositions among the parent bodies of most meteorites is unclear, and is important for constraining the Earth's origin and the likely composition of a Moon-forming impactor.

In conclusion, high precision chemical and isotopic analyses of returned lunar samples have shaken the foundations of the paradigm of lunar formation by a giant impact. However, a multitude of new concepts have emerged whose details and implications still need to be evaluated. This, together with increasing prospects for further lunar exploration in the near-term, makes this a truly exciting time for lunar origin science.

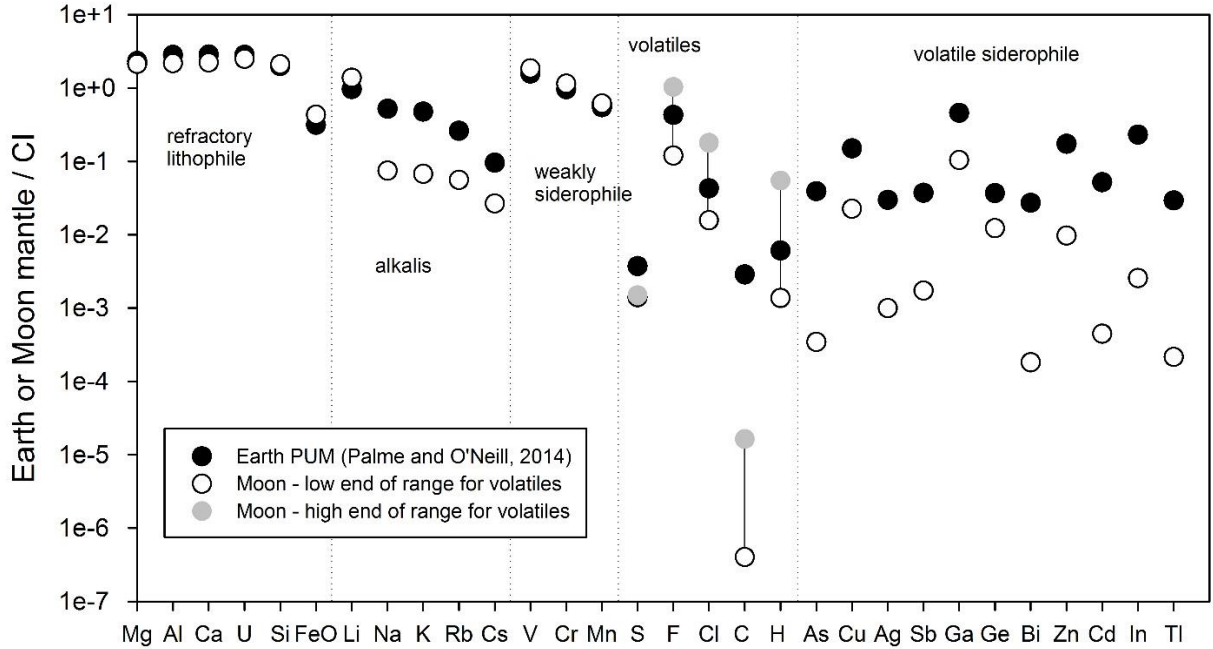

**Figure 1.** Abundance of elements in the bulk silicate Earth/primitive upper mantle (PUM; solid circles; Palme and O'Neill 2014) and the bulk silicate Moon (light and open circles; see references below) normalized to abundances in CI chondrites (Newsom 1995). Groups of elements are the refractory lithophile (Mg, Al, Ca, U, Si, FeO), alkalis (Li, Na, K, Rb, Cs), weakly siderophile (V, Cr, Mn), classic volatiles (S, F, Cl, C H), and the volatile siderophile elements (VSE, ordered by increasing volatility as gauged by their 50% condensation temperature from Lodders 2003). Lunar estimates are from Taylor (1982), Taylor and Wieczorek (2014), Righter et al. (2011, 2018a), Hauri et al. (2015) and McCubbin et al. (2015). Li estimates for the lunar mantle (2 ppm) are calculated assuming a bulk D(Li) for mantle melting of 0.2 (based on experiments of Herd et al. 2004), 7 ppm Li in a lunar mantle melt (with degree of melting F=0.1. For the volatiles (for all except sulfur) a high and low end of range are given. Fluorine ranges from 4.5 (Hauri et al. 2015) to 60 ppm (McCubbin et al. 2015). Chlorine ranges from 0.142 ppm (Hauri et al. 2015) to 123 ppm (McCubbin et al. 2015). Carbon ranges from 0.014 to 0.57 ppm (McCubbin et al. 2015), and $H_2O$ ranges from 3.1 (McCubbin et al. 2015) to 292 ppm (Hauri et al. 2015). Sulfur estimates from McCubbin et al. (2015) and Hauri et al. (2015) are both 78.9 and as such no range is indicated here. Volatile siderophile elements show lunar depletions that are similar to the alkalis (e.g., Ga, Cu) or even higher (As, Ag, Sb, Ge, Bi, Zn, Cd, In, Tl) that may be due to core formation or later magmatic degassing (see text for additional discussion).

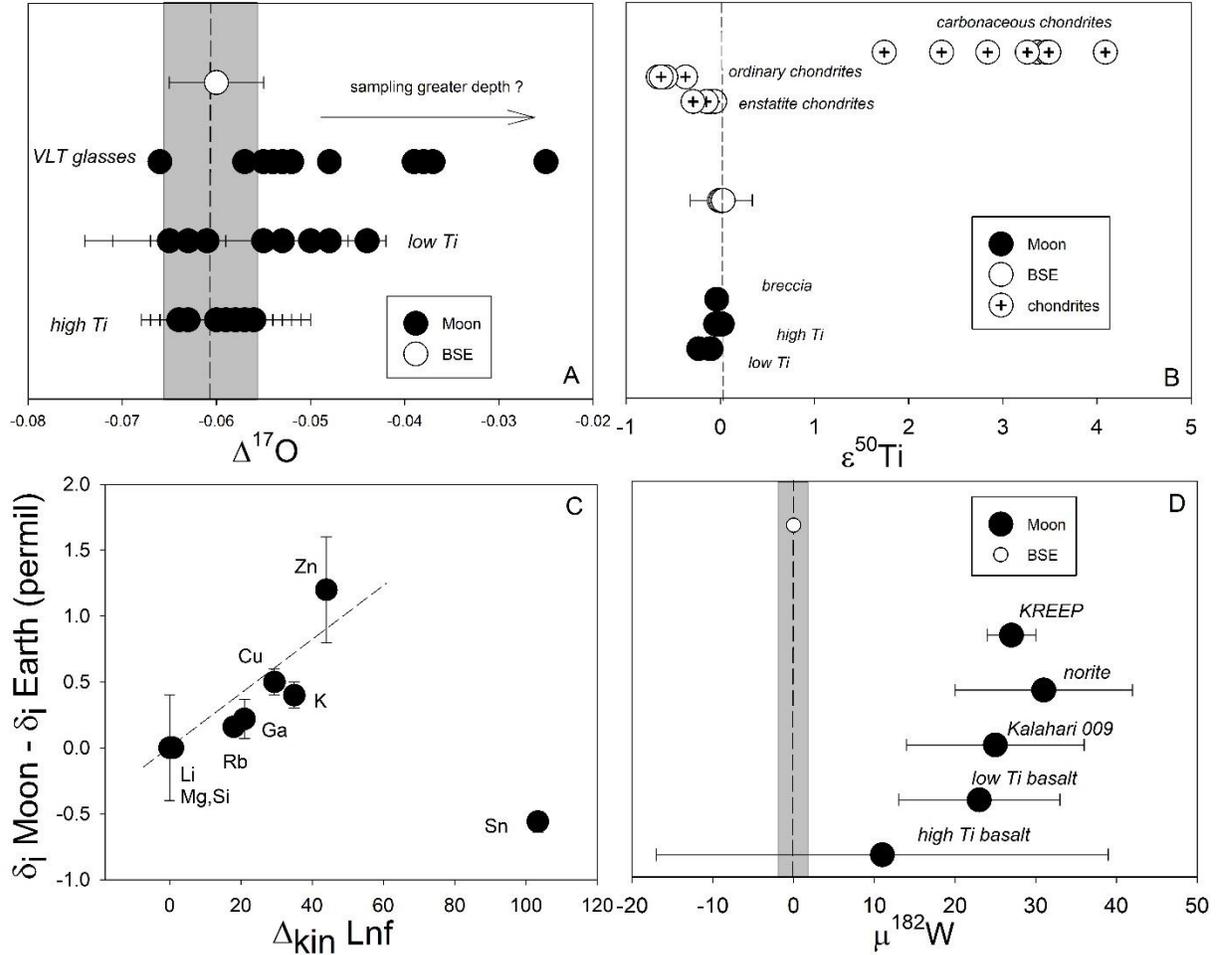

**Figure 2.** (A) $\Delta^{17}O$ isotopic values for lunar and terrestrial samples (from Cano et al. 2020). (B) $\varepsilon^{50}Ti$ (= $[(^{50}Ti/^{47}Ti)_{sample}/(^{50}Ti/^{47}Ti)_{rutile} - 1] \times 10^4$) for lunar and terrestrial samples, and carbonaceous, enstatite, and ordinary chondrites (from Zhang et al. 2012). (C) Relationship between stable isotopic differences between Earth and Moon ($\delta_i$ Moon - $\delta_i$ Earth), and a parameter controlling isotopic fractionation during evaporation in a kinetic regime ($\Delta_{kin} \ln f$, with $\Delta_{kin} = [(m_1/m_2)^\beta-1]*1000$ and $f$ = degree of depletion in the bulk Moon relative to the bulk Earth) (modified from Nie and Dauphas 2019). If elements are lost by evaporation under a kinetic regime in a medium similarly undersaturated for all elements, one would expect to find a straight line, whose slope gives the degree of undersaturation in the medium. Using updated depletion levels from Righter (2019), we calculate a slightly lower undersaturation of 0.98 compared to that calculated by Nie and Dauphas (2019) of 0.99. Given the lack of predictive power of existing models of MVE depletion, such a subtle difference in degree of undersaturation during evaporation is largely inconsequential. Available Sn isotope data however cannot be explained by kinetic isotopic fractionation and if they are representative of the bulk Moon would require equilibrium isotopic fractionation (Bourdon and Fitoussi 2020). Further work is therefore needed to evaluate whether loss of moderately volatile elements from the Moon occurred under kinetic or equilibrium processes, or a combination of the two. Data sources: Wang et al. (2019), Nie and Dauphas (2019)

and references therein; f calculated for Sn, Zn, Cu, and Ga using compilation of Righter (2019). (D) W isotopic measurements from Kruijer and Kleine (2017) illustrate the current (i.e., post late accretion) ~26 ppm $^{182}$W difference between BSE and BSM. Here $\mu^{182}$W is the parts per $10^6$ variation of $^{182}$W/$^{184}$W compared to a terrestrial standard.

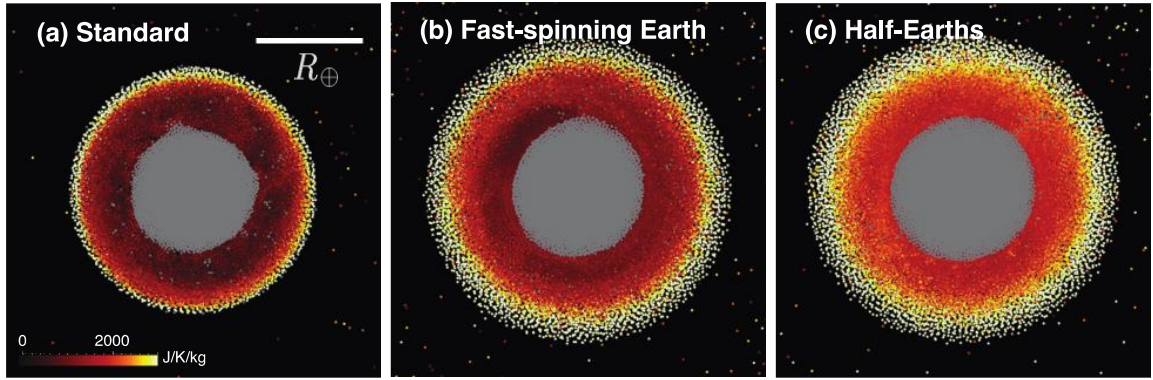

**Figure 3**. Thermal state of the Earth's mantle after the Moon-forming impact. The red-yellow color shows the impact-induced entropy increase and grey shows iron (after Nakajima and Stevenson 2015). The mantle is more shock-heated in the models (b) and (c).

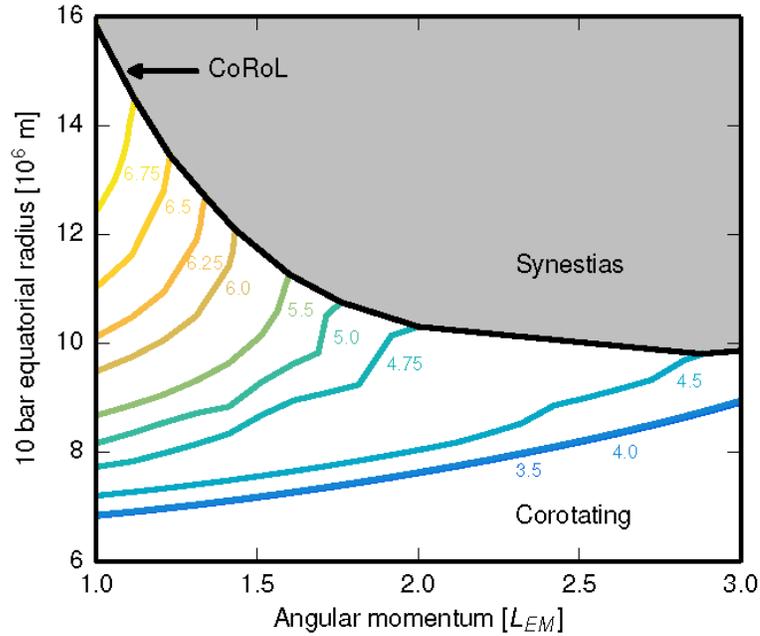

**Figure 4.** As the internal energy and angular momentum of a planet increases, the body crosses the corotation limit (black line) and becomes a synestia. Internal energy is represented by the specific entropy of an adiabatic temperature profile. Each colored line is the range of equatorial radii of bodies with stratified thermal profiles calculated using the HERCULES code (Lock and Stewart 2017). The upper silicate layer, defined by the 25 wt% of the total silicate at the lowest pressures, is assumed to be isentropic until it intersects the liquid-vapor phase boundary at lower pressures, at which point it follows the vapor side of the boundary. The specific entropy of the isentropic section for the upper silicate layer is indicated by the line color, with values shown in units of kJ K$^{-1}$ kg$^{-1}$. For reference, the critical point value is 5.4 kJ K$^{-1}$ kg$^{-1}$ for the forsterite equation of state, and entropies above this value correspond to mostly vaporized or supercritical states. The lower silicate layer has a specific entropy of 4 kJ K$^{-1}$ kg$^{-1}$, corresponding to an adiabat through liquid forsterite typical of calculated post-impact structures (Fig. 5).

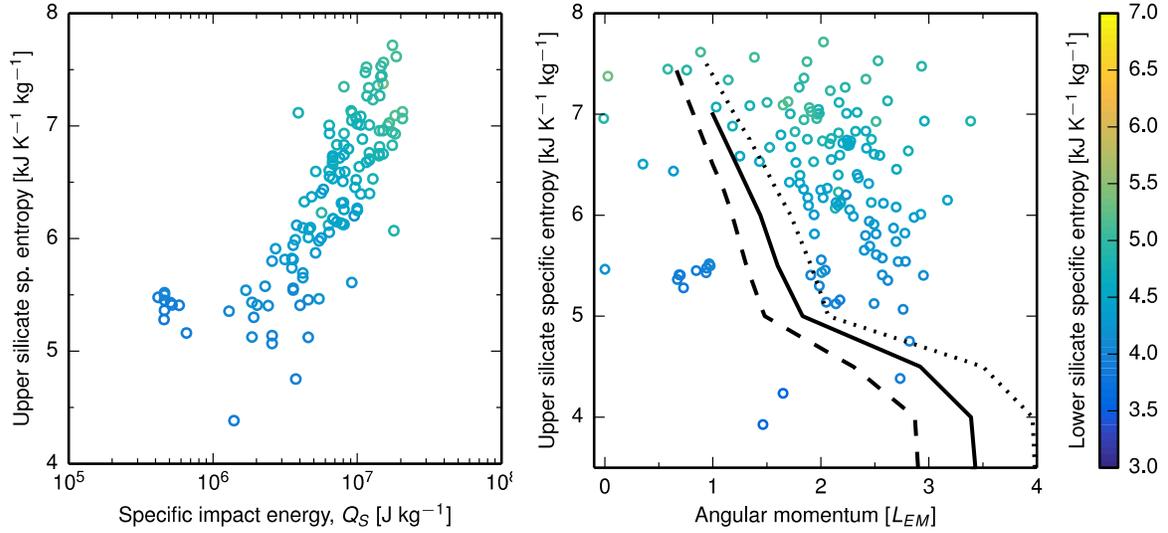

**Figure 5.** Giant impacts can produce thermally stratified bodies that exceed the corotation limit. Points are derived from the giant impact database in Lock and Stewart (2017) that includes a wide range of impact configurations with final bound masses between 0.9 and $1.1 M_\oplus$. **A.** The specific entropy of the upper 25 wt.% silicate layers increases with the specific energy of the impact, $Q_S$ (center of mass kinetic energy divided by the total mass with a geometric adjustment). The outer silicate layers are largely vaporized with specific entropies greater than the critical point value of 5.4 kJ K$^{-1}$ kg$^{-1}$ for the forsterite equation of state. **B.** The production of synestias as a function of the post-impact angular momentum. The lines denote the corotation limit for Earth-like composition bodies of 0.9 (dashed), 1.0 (solid) and 1.1 (dotted) $M_\oplus$ with the same thermal structure as the bodies used in Figure 4. The colors indicate the average specific entropy of the corresponding lower 75 wt% of silicates. Figure modified from Lock and Stewart (2017).

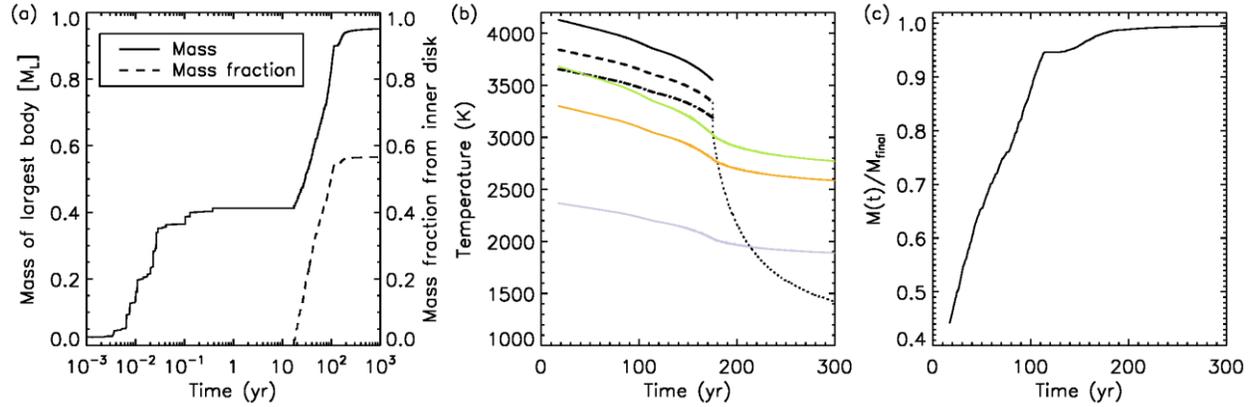

**Figure 6**. Lunar accretion model and implications for lunar volatile content (figures modified from those in Salmon and Canup 2012 and Canup et al. 2015). (a) Mass of the growing moon (solid line) and fraction of its mass derived from the Roche-interior disk (dashed line), as a function of time, for a sample simulation. Inner-disk material is accreted by the Moon only during a protracted phase that starts ~10 years into the accretion process, and lasts for ~100 years. (b) Estimates of midplane and condensation temperatures at the Roche limit vs. time for the simulation shown in panel (a); note change to a linear *x*-axis scaling. Black solid, dashed, and dot-dashed lines show the midplane temperature for different assumed gas mass fractions at the midplane using the disk model of Ward (2012). The dotted curve considers a radiatively cooling disk after the inner silicate vapor has condensed. Colored curves show estimated 50% condensation temperatures ($T_{50}$) for key volatiles: K (green), Na (orange), and Zn (grey). (c) The Moon's mass scaled to its final mass ($M_{final}$). The Moon accretes only about one percent of its mass after the temperature of clumps falls substantially below $T_{50}$ for potassium. After this time, volatile-rich clumps near the Roche limit are instead scattered back towards the Earth by the Moon.

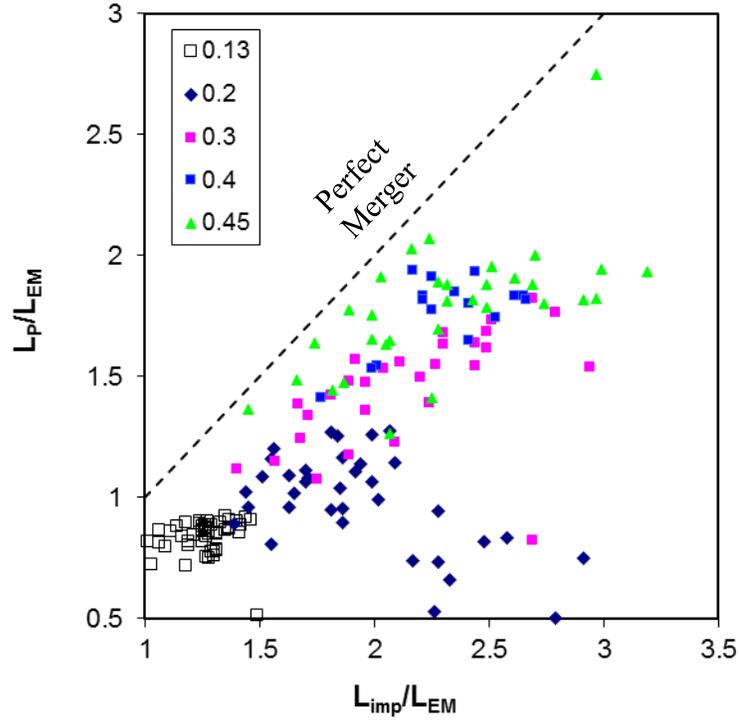

**Figure 7**. The angular momentum in an Earth-mass planet's rotation, $L_P$, delivered by a single impact with angular momentum $L_{imp}$, based on SPH simulations of impacts across a broad range of impact angles and speeds that consider nonrotating targets (Canup 2004a, 2012). Impactor-to-total mass ratio ($\gamma$) is indicated by color in the legend. The dashed line shows the limit of a perfect merger in which all of the impact angular momentum is retained in the planet's rotation; actual values fall below this line due to material that escapes or goes into orbit around the planet. For planets whose moments of inertia vary with planet mass and radius as $M_P R_P^2$, the rotational angular momentum for a fixed angular rotation rate varies as $M_P^{5/3}$. Here we scale the planet angular momenta by a factor $(M_P/M_\oplus)^{5/3}$ to produce results appropriate for an Earth-mass planet and to remove the effect of differences in final planet mass across different simulations (nearly all cases shown have $M_P$ within 10% of $M_\oplus$).